\documentclass[pre,onecolumn,notitlepage,longbibliography,amsmath,amssymb,floats,superscriptaddress,nofootinbib,10pt]{revtex4-2}
\usepackage{amsmath}
\usepackage[colorlinks=true,linkcolor=blue,citecolor=blue,urlcolor=blue]{hyperref}
\usepackage[normalem]{ulem}
\usepackage{soul}
\usepackage{graphicx}
\newcommand{\nn}{\nonumber\\}
\newcommand{\lb}{\left(}
\newcommand{\rb}{\right)}

\usepackage[dvipsnames]{xcolor}

\definecolor{ored}{RGB}{0,0,255}

\begin{document}

\title{Non-relativistic transport from frame-indifferent kinetic theory}
\begin{abstract}
This paper explores the application of Newton-Cartan geometry to the kinetic theory of gases that includes non-relativistic gravitational effects and the principle of general covariance. Starting with an introduction to the basics of Newton-Cartan geometry, we examine the motion of point particles within this framework, leading to a detailed analysis of kinetic theory and the derivation of conservation equations. The equilibrium distribution function is explored, and the example of a rotating gas in a gravitational field is discussed. Further, we develop covariant hydrodynamic equations and extend our analysis through a gradient expansion approach to assess first-order constitutive relations for rotating gases. Finally, we address the frame-dependence paradox, presenting a novel resolution that addresses apparent discrepancies. Our construction resolves a fifty-year-old debate about the frame-indifferent formulation of kinetic theory. The resolution is presented in a modern, symmetry-based approach.
\end{abstract}

\author{Paweł Matus}
\affiliation{Max Planck Institute for the Physics of Complex Systems and W\"urzburg-Dresden Cluster of Excellence ct.qmat, 01187 Dresden, Germany}

\author{Rajesh Biswas}
\affiliation{Institute of Theoretical Physics, Wroc\l{}aw  University  of  Science  and  Technology,  50-370  Wroc\l{}aw,  Poland}

\author{Piotr Sur\'{o}wka}
\affiliation{Institute of Theoretical Physics, Wroc\l{}aw  University  of  Science  and  Technology,  50-370  Wroc\l{}aw,  Poland}

\author{Francisco Pe\~na-Ben\'itez}
\affiliation{Institute of Theoretical Physics, Wroc\l{}aw  University  of  Science  and  Technology,  50-370  Wroc\l{}aw,  Poland}

\maketitle
\flushbottom

\section{Introduction}

Kinetic theory is an indispensable tool from statistical physics toolbox that sheds light on a broad range of phenomena including classical gasses, active matter, ultrapure metals, quantum topological materials, light-matter interactions and cosmological evolution, to name a few examples~\cite{chapman1990mathematical,lifshitz1981physical,dorfman2021contemporary,vereshchagin2017relativistic,Romatschke:2017ejr}. Despite its fundamental role and long history there are aspects of kinetic theory that are only partially understood and crucial to clarify the interplay between transport, topology and geometry. Two such aspects are how to formulate Galilean-invariant kinetic theory on curved backgrounds and non-inertial frames. In other words, one would like to construct equations that are in accord with the principle of covariance under general coordinate transformations. 

This seemingly simple task has created various confusions and debates throughout the development of kinetic theory. In 1972, I. M{\"u}ller in Ref. \cite{Muller1972} presented calculations that seemed to prove that stress and heat fluxes in a gas computed from kinetic theory do not take consistent frame-covariant expressions. A lively debate followed with various erroneous arguments proposed both for and against the frame invariance principle \cite{Edelen1973,Wang1975,Wang1976,Soderholm1976,Truesdell1976,Speziale1981,Murdoch1982,Lebon1985}. Eventually, the authors of Refs. \cite{Woods1983, Band1984,Eu1985} independently proposed that the solution lies in calculating constitutive relations in the frame \textit{co-rotating} with the fluid. Simultaneously, Refs. \cite{Lianis1979,Ryskin1985} recognized that the paradox can be avoided in a relativistic-like four-dimensional treatment. Despite partial successes in formulating the kinetic theory in a covariant way, the fundamental reasons behind the problems have remained obscure and the debate continues up to the present date \cite{Matolcsi1986,Speziale1986,Ryskin1988,Lebon1988,Matolcsi1996,Sadiki1996,Muschik2002,Murdoch2003,Liu2004,Liu2005,Murdoch2005,Barbera2006,Matolcsi2006,Matolcsi2007,Muschik2008,Frewer2009,Romano2018,Soldatos2020}. For a more comprehensive summary of the debate, see Ref. \cite{Frewer2009}.

 In contrast, developments in hydrodynamics have successfully implemented a fully covariant framework using the Newton-Cartan (NC) formalism~\cite{Jensen:2014aia,Jensen:2014wha}. The NC theory, originally introduced to geometrically formalize non-relativistic Newtonian gravity~\cite{Cartan:1923zea,Cartan:1924yea}, can be seen as a $1/c^2$ expansion of general relativity \cite{Hansen:2020pqs,Vanden2017}.  Its covariant nature not only upholds the principle of Galilean invariance but also facilitates the study of non-relativistic motion on curved spacetimes. However, despite the potential applications of NC geometry in kinetic theory, there has been no significant attempt to integrate it, even though it is a natural framework to address issues related to frame-invariance. Previous studies of kinetic theories on manifolds have generally failed to preserve the symmetries inherent to NC geometries \cite{love_boltzmann_2011}.

The development of kinetic theory for rotating fluids on curved surfaces, particularly through the derivation of odd viscosity from microscopic dynamics, opens new avenues in the study of fluid behavior under rotation. Odd viscosity is a non-dissipative term that emerges in the equations governing rotating fluids, reflecting their inherent lack of time-reversal symmetry—a characteristic not accounted for in classical fluid dynamics. By applying these insights to macroscopic fluid models, such as rotating shallow water systems, researchers can incorporate odd viscosity to regularize otherwise pathological behaviors in the spectral properties of fluid waves at high wavenumbers \cite{tauber_bulk-interface_2019,tauber_anomalous_2020,graf_topology_2021}. Our computation generalizes previous non-covariant computations of odd transport based on rotating reference frames \cite{nakagawa_kinetic_1956,Aung2024}.

In addition to the direct applications of our formalism to the microscopic computations of transport phenomena in rotating classical fluids, there are complex systems such as rotating quantum fluids that necessitate a more generalized approach to kinetic theory. Our methods provide a foundational framework for exploring these systems. Examples of such systems include multi-Weyl semimetals, where the chiral vortical effect becomes prominent in a system breaking Lorentz symmetries \cite{dantas_non-abelian_2020}, rotating superfluids at finite temperatures \cite{sonin_dynamics_2015,svistunov_superfluid_2015}, characterized by a blend of a normal fluid component and a quantum condensate (for the kinetic theory description see \cite{hohenberg_microscopic_2000}). These cases highlight the necessity and potential of our methods to facilitate deeper insights into the microscopic properties of quantum materials under rotation \cite{frohlich_gauge_2023}.

This manuscript is organized as follows. In Sec. \ref{sec:NC_geometry} we review basic facts about the NC geometry. In Sec. \ref{sec:point} we study the single-particle Lagrangian and the single-particle motion. In Sec. \ref{sec:kinetic} we formulate the Boltzmann equation and introduce the collision operator, and proceed to derive conservation equations for the particle number, momentum and energy, as well as the general expression for an equilibrium distribution function. In Sec. \ref{sec:rotate} we put the formalism introduced up to this point to a practical use by calculating the equilibrium distribution of a rotating gas in three dimensions. In Sec. \ref{sec:hydro} we take a step towards non-equilibrium physics by formulating covariant equations of hydrodynamics and calculating transport coefficients in two- and three-dimensional gases under the action of magnetic field and/or the Coriolis force. Finally, in Sec. \ref{sec:invariance} we come back to the issue of the frame-invariance controversy, explaining how the problem can be understood and resolved using the NC kinetic theory. We also discuss briefly two especially interesting former approaches to the issue \cite{Matolcsi1986,Matolcsi1996,Frewer2009}.


\section{Basics of Newton-Cartan geometry}
\label{sec:NC_geometry}

Newton-Cartan (NC) spacetimes can be constructed from a non-relativistic limit of pseudo-Riemannian geometries. Usually, this limit is taken either by an explicit expansion in $1/c^2$ \cite{Hansen:2020pqs,Vanden2017} or by a so-called light-cone reduction \cite{Duval:1984cj,Son:2008ye,Banerjee:2015hra}. However, in this paper, we will follow a more axiomatic perspective in the spirit of \cite{Jensen:2014aia,Jensen:2014ama}. 

A NC spacetime is a $(d+1)$-manifold with coordinates $x^\mu$ with $\mu=0,1,\ldots d$ and:
\begin{itemize}
    \item a 1-form $\tau_\mu$ usually called the clock form or time metric, such that $\tau_\mu (x_2^\mu - x_1^\mu)$ measures the passage of time between two infinitesimally close points $x_1^\mu$ and $x_2^\nu$; 
    \item a degenerate symmetric tensor $h^{\mu\nu}$, with signature $(0,1,\ldots,1)$, that satisfies $\tau_\mu h^{\mu\nu}=0$, interpreted as inverse spatial metric; 
    \item a covariant derivative  $\nabla_\mu$ that acts on a tensor $T_\alpha^{~\beta}$ as 
    \begin{equation}
    \nabla_\mu T_\alpha^{~\beta} = \partial_\mu T_\alpha^{~\beta} +\Gamma^\beta_{\lambda\mu}T_\alpha^{~\lambda}-\Gamma^\lambda_{\alpha\mu}T_\lambda^{~\beta}.
    \end{equation}
    As in general relativity, we require it to be compatible with the geometric data, $\nabla_\mu\tau_\nu=\nabla_\mu h^{\nu\rho}=0$.
\end{itemize}

Throughout the paper, we raise indices using the inverse spatial metric, eg. $T^{\alpha\beta}=h^{\alpha\gamma}T_\gamma^{~\beta}$. Note that since $h^{\mu\nu}$ is degenerate, raising an index is not invertible; therefore, the distinction between upper-index and lower-index components of tensors is more significant than in relativistic physics.
We assume that the coordinates, the time metric, and the spatial metric have units $[x^\mu]=\mathrm{length}$, $[\tau_\mu]=\mathrm{time}/\mathrm{length}$, and $[h^{\mu\nu}]=1$, respectively.

Now let us introduce a family of observers labelled by $\psi$ with an associated velocity field $(v_\psi)^\mu$, normalized such that $\tau_\mu(v_\psi)^\mu=1$. Any such velocity field can be expressed as a \textit{Milne boost} of a chosen reference velocity
\begin{equation}
    (v_\psi)^\mu = v^\mu + h^{\mu\nu}\psi_\nu\,,\label{eq:redf-v}
\end{equation}
with $ v^\mu$ an arbitrary reference velocity satisfying $\tau_\mu v^\mu=1$. The reference velocity $v^\mu$ makes it possible to define a metric $h_{\mu\nu}$ satisfying 
\begin{equation}
    h_{\mu\nu}v^\nu=0\,,\quad h^{\mu\nu}h_{\nu\rho}+v^\mu\tau_\rho =\delta^\mu_\rho\,.
\end{equation}
Milne boosts express a \emph{gauge freedom} in the choice of geometric data in NC geometry. Specifically, they reflect the freedom to redefine the reference velocity without altering the underlying physical structure. This gauge freedom is implemented via transformations given by Eq. \eqref{eq:redf-v}, and leads to a corresponding transformation of the spatial metric. As a result, the spatial metric \( h_{\mu\nu} \) becomes \emph{observer-dependent}, transforming under a Milne boost as:
\begin{equation}
    (h_\psi)_{\mu\nu} = h_{\mu\nu} - \left( \tau_\mu P^\alpha_{\nu} + \tau_\nu P^\alpha_{\mu} \right) \psi_\alpha + \psi^2 \tau_\mu \tau_\nu\,,
    \label{eq:redf-h}
\end{equation}
where \( P^\mu_{\nu} = \delta^\mu_{\nu} - v^\mu \tau_\nu \) is the spatial projector and \( \psi^2 = h^{\alpha\beta} \psi_\alpha \psi_\beta \). This structure captures the fact that Newton--Cartan geometry is not fixed by a unique choice of \( v^\mu \), but instead allows for an entire gauge orbit of related geometric data connected by Milne boosts. Requesting Milne boost invariance of the theory can be interpreted as the independence of physical phenomena from the observer describing them.

As shown in~\cite{Jensen:2014aia}, the affine connection \( \Gamma^\mu_{\alpha\beta} \) must transform accordingly to preserve Milne invariance. With these ingredients we find that the connection $\Gamma^\mu_{\alpha\beta}$ has the general form~\cite{Jensen:2014aia}\footnote{We will use round brackets to denote symmetrization $A_{(\mu}B_{\nu)}=\frac{1}{2}\lb A_\mu B_\nu + A_\nu B_\mu\rb$, and square brackets for antisymmetrization $A_{[\mu}B_{\nu]}=\frac{1}{2}\lb A_\mu B_\nu - A_\nu B_\mu\rb$ respectively. } 
\begin{align}
    \Gamma^{\mu}_{\alpha\beta}=v^\mu \partial_{\beta}\tau_\alpha + \frac{1}{2}h^{\mu\sigma}\lb \partial_\alpha h_{\beta\sigma} + \partial_{\beta}h_{\alpha\sigma} - \partial_\sigma h_{\alpha\beta}\rb+ h^{\mu\sigma}\tau_{(\alpha}F_{\beta)\sigma}\,,
    \label{eq:NC-connection}
\end{align}
with $F_{\mu\nu}$ an antisymmetric tensor with units $[F_{\mu\nu}]=\mathrm{time}^{-1}$. A straightforward but tedious computation reveals that the geodesic acceleration $a_v^\mu\equiv v^\nu \nabla_\nu v^\mu$ and vorticity $\omega_v^{\mu\nu}\equiv 2\nabla^{[\mu} v^{\nu]}$ are given by
\begin{equation}
    a_v^\mu=-F^\mu\,_\sigma v^\sigma\,,~~~~\omega_v^{\mu\nu} = F^{\mu\nu}\,,
    \label{eq:acceleration_curl}
\end{equation}
where the indices are raised with  $h^{\mu\nu}$. The connection defined in Eq.~(\ref{eq:NC-connection}) is in general observer-dependent, but if we introduce a frame-dependent gauge field $m_\mu$ such that $F_{\mu\nu}=\partial_\mu m_\nu - \partial_\nu m_\mu$ and postulate the transformation
\begin{align}
    (m_\psi)_\mu=m_\mu + P^\nu_{\mu}\psi_\nu - \frac{\psi^2}{2}\tau_\mu +\partial_\mu\Lambda\,,
    \label{eq:redf-m}
\end{align}
with $\Lambda$ the gauge parameter, the connection transforms as
\begin{equation}
    \lb\Gamma_\psi\rb^\mu_{\alpha\beta}=\Gamma^\mu_{\alpha\beta}-\frac{1}{2}h^{\mu\sigma}\lb\psi_\sigma H_{\alpha\beta}+2\psi_{\rho}P^{\rho}_{(\alpha}H_{\beta)\sigma}- \psi^2 \tau_{(\alpha}H_{\beta)\sigma}\rb\,,
    \label{eq:gamma_milne}
\end{equation}
where
\begin{equation}
    H_{\mu\nu}\equiv\lb d\tau\rb_{\mu\nu}=\partial_{\mu}\tau_{\nu}-\partial_\nu \tau_\mu\,.
    \label{eq:h_munu}
\end{equation}
Thus, the connection $\Gamma^\mu_{\alpha\beta}$ becomes observer-invariant if $d \tau=0$, which is always the case when there exists a globally defined time function such that $\tau=d t$, and which is furthermore equivalent to setting the torsion $\Gamma^{\mu}_{[\alpha\beta]}$ to zero.

Finally, a volume form satisfying the  Milne- and gauge-invariance requirement can be defined after introducing the non-degenerate tensor $G_{\mu\nu}=c^2\tau_\mu \tau_\nu + h_{\mu\nu}$ that allows us to define the volume form\footnote{To construct $G_{\mu\nu}$ it is necessary to have a velocity scale $c$. This parameter a priori does not have to be the speed of light. However, as we will see below, observables will not depend on such a constant; therefore, we can fix it to be the speed of light.}
\begin{equation}
    \omega_{d+1} = \frac{1}{(d+1)!}\epsilon_{\mu_0\ldots\mu_d} dx^{\mu_0\ldots\mu_d}=\sqrt{G}d^{d+1}x\,,\label{eq:VolNCspace}
\end{equation}
with $G$ the determinant of $G_{\mu\nu}$, and the fully antisymmetric tensor satisfying $\epsilon_{01\ldots d}=\sqrt{G}$.
\newline
\subsection{Flat NC geometry, Milne boosts and Galilean transformations}\label{sec:FlatNC}
From the perspective of the Galilean group, the presence of the gauge field $m_\mu$ is natural since mass is a conserved charge for non-relativistic matter, and therefore we can expect the presence of a gauge field that couples to the mass current. However, the Milne transformations are less intuitive and they do not have a counterpart for relativistic systems. As we will discuss below, Milne boosts are a necessary ingredient to obtain the Galilean group as the isometry group of the flat NC geometry.
\newline
Let us first define the set of infinitesimal coordinate reparametrization, Milne boost, and gauge transformation as $\chi=(\xi^\mu\partial_\mu,\psi_\mu dx^\mu,\Lambda)$. The variation $\delta_\chi$ of the NC data reads
\begin{align}
    \delta_\chi \tau_\mu=&\pounds_\xi\tau_\mu\,,
    \label{eq:var_of_tau}
    \\
    \delta_\chi h^{\mu\nu}=&\pounds_\xi h^{\mu\nu}\,,
    \label{eq:var_of_h}
    \\
    \delta_\chi v^\mu = & \pounds_\xi v^\mu +h^{\mu\nu}\psi_\nu\,, 
    \label{eq:var_of_v}
    \\
    \delta_\chi m_\mu = & \pounds_\xi m_\mu +P^\nu_\mu\psi_\nu +\partial_\mu \Lambda\,,
    \label{eq:var_of_A}
\end{align}
where $\pounds_\xi$ denotes the Lie derivative along $\xi^\mu$.
These transformations form an algebra as $\left[\delta_{\chi_{1}},\delta_{\chi_{2}}\right]=\delta_{\chi_{[12]}}$, where $\chi_1 = \left(\xi^\mu_1,\psi_\mu^1, \Lambda_1\right)$ and $\chi_2 = \left(\xi^\mu_2,\psi_\mu^2, \Lambda_2\right)$ are two sets of symmetric parameters, and $\chi_{[12]}=\left(\xi^\mu_{[12]},\psi_\mu^{[12]}, \Lambda_{[12]}\right)$ with
\begin{align}
    \xi^\mu_{[12]} &= \pounds_{\xi_1}\xi_2\,,\\
    \psi_\mu^{[12]} &=\pounds_{\xi_1}\psi_\mu^2 - \pounds_{\xi_2}\psi_\mu^1\,,\\
    \Lambda_{[12]} &= \pounds_{\xi_1}\Lambda_2 - \pounds_{\xi_2}\Lambda_1\,.
\end{align}
In particular, the flat NC geometry is given by
\begin{equation}\label{eq:flatNC}
    \tau_\mu = c^{-1}\delta_\mu^0\,,\quad v^\mu = c\delta^\mu_0\,,\quad h^{\mu\nu} =  \delta^\mu_i\delta^\nu_j \delta^{ij}\,,\quad m_\mu=0\,,
\end{equation}
and its isometry group is the set of transformations $\delta_{\chi}$ leaving the fields invariant. In particular, the set of Killing generators is
\begin{align}
    H &= \left(-c\partial_0,0,0\right) \,,\qquad M = \left(0,0,1\right) \,,\\
    P_i &= \left( - \partial_i,0,0\right)\,, ~\qquad K_i = \left(-c^{-1}x^0\partial_i , -dx^i, x^i \right) \,,\\
    R_{ij} &= \left(-x^i\partial_j + x^j\partial_i, 0 ,0\right)\,,
\end{align}
satisfying the commutation relations of the so-called Bargmann algebra\footnote{This is an extension of the Galilean algebra with a central element $ M $, associated with mass.}
\begin{equation}
\begin{split}
    &[R_{ij}, R_{kl}] = \delta_{ik} R_{jl} - \delta_{il} R_{jk} + \delta_{jl} R_{ik} - \delta_{jk} R_{il}\,, \\
    &[R_{ij}, P_k] = \delta_{ik} P_j - \delta_{jk} P_i\,, \qquad
[R_{ij}, K_k] = \delta_{ik} K_j - \delta_{jk} K_i\,, \\
    &[K_i, H] =  P_i\,, \qquad \qquad\qquad\quad~
    [K_i, P_j] =  \delta_{ij} M \,.
    \label{eq:Bargman_Algebra} 
\end{split}
\end{equation}
The generators correspond to time translations $( H )$, spatial translations $( P_i )$, rotations $( R_{ij} )$, Galilean boosts $( K_i )$, and the central charge $( M )$. We note that the index on the generators serves as a label and does not correspond to a spatial component. Such a construction allows one to have the proper field content that couples to the energy, momentum, and mass currents. Moreover, we emphasize that without the Milne boosts the algebra in Eqs. \eqref{eq:Bargman_Algebra} would not close. 

\subsection{Thermal NC spacetime}
\label{sec:TNC}

In \cite{Jensen:2014ama} NC spacetimes were generalized by adding to the manifold an extra timelike vector field $u^\mu$, 
normalized as $u^\mu\tau_\mu=1$. As we will see later, 
at finite density such a field can always be defined
, in which case $u^\mu$ describes the velocity of the gas. For such systems 
we could argue that the ``Milne symmetry" is spontaneously broken since $u^\mu$ introduces a distinguished Milne frame with velocity $v_u^\mu=u^\mu$. Notice that this frame is connected to an arbitrary one via a Milne boost $v^\mu-v_u^\mu = h^{\mu\nu}\psi_\nu$. The presence of $u^\mu$ allows us to define the spatial metric $g_{\mu\nu}$ and $U(1)$ gauge field $A_\mu$
\begin{align}
 \label{eq:hU}    g_{\mu\nu} &= h_{\mu\nu} - u_\mu\tau_\nu - u_\nu\tau_\mu + u^2\tau_\mu\tau_\nu\,,\\
     A_{\mu} &= m_\mu + u_\mu -\frac{u^2}{2}\tau_\mu \,,
    \label{eq:A_inv}
\end{align}
where $u_\mu=h_{\mu\nu}u^\nu$ and $u^2=h_{\mu\nu}u^\mu u^{\nu}$.
The objects $g_{\mu\nu}$ and $A_{\mu}$ are precisely equal to $h_{\mu\nu}$ and $m_{\mu}$ evaluated in the distinguished frame $v_u^\mu$. A straightforward calculation shows that even though the definitions~(\ref{eq:hU}) and~(\ref{eq:A_inv}) involve $h_{\mu\nu}$ and $m_\mu$, $g_{\mu\nu}$ and $A_{\mu}$ do not transform under Milne boosts as described by Eqs.~(\ref{eq:redf-h}),~(\ref{eq:redf-m}) and are therefore defined unambiguously. We furthermore introduce the distinguished spatial projector and an affine connection corresponding to $v_u^\mu$ as, respectively,  \cite{Jensen:2014ama}
\begin{align}
    \bar{P}^\mu_\nu &= \delta^\mu_\nu - u^\mu \tau_\nu =  h^{\mu\sigma}g_{\nu\sigma}\,,
    \label{eq:P_inv}\\
    \bar{\Gamma}^{\mu}_{\alpha\beta} & =u^\mu \partial_{\beta}\tau_\alpha + \frac{1}{2}h^{\mu\sigma}\lb \partial_\alpha g_{\beta\sigma} + \partial_{\beta}g_{\alpha\sigma} - \partial_\sigma g_{\alpha\beta}\rb+ h^{\mu\sigma}\tau_{(\alpha}f_{\beta)\sigma}\,,
    \label{eq:NC-connection2}
\end{align}
with
\begin{equation}
    f=dA\,.
\end{equation}
A comparison with Eq.~(\ref{eq:gamma_milne}) reveals that $\bar{\Gamma}^{\mu}_{\alpha\beta}=\Gamma^{\mu}_{\alpha\beta}$ in the torsionless case. Using the connection $\bar{\Gamma}^{\mu}_{\alpha\beta}$, which is both gauge- and Milne-invariant, the covariant derivative of the velocity field can be decomposed as
\begin{equation}
    \bar\nabla_\mu u^\nu = \tau_\mu a^\nu + \frac{1}{2}g_{\mu\alpha}\omega^{\alpha\nu} + g_{\mu\alpha}\sigma^{\alpha\nu} + \frac{1}{d}\bar P_\mu^\nu\theta\,,
    \label{eq:nabla_u_decompose}
\end{equation}
with the acceleration, vorticity, shear tensor, and compression defined as
\begin{align}
   a^\mu &\equiv u^\nu\bar\nabla_\nu u^\mu = - f^\mu\,_\nu u^\nu\,,\\ 
   \omega^{\mu\nu} &\equiv \bar\nabla^\mu u^\nu - \bar\nabla^\nu u^\mu = f^{\mu\nu}\,,\\
   \sigma^{\mu\nu} &\equiv \bar\nabla^{\langle \mu}u^{\nu\rangle}\,,\\
   \theta &\equiv \bar\nabla_\mu u^\mu\,, \label{eq:def_theta}
\end{align}
where the angle brackets denote the traceless symmetric part of a tensor, e.g.,
\begin{equation}
    A^{\langle\alpha\beta\rangle} = A^{(\alpha\beta)}-\frac{1}{d}h^{\alpha\beta}g_{\gamma\delta}A^{\gamma\delta}.
\end{equation}

The new structure added to the NC manifold may seem unnecessary. However, in what follows we will prove that the existence of the notion of thermal equilibrium is only possible on this class of NC spaces. Therefore, we will call them Thermal Newton-Cartan (TNC) manifolds.

\section{Motion of a point particle}
\label{sec:point}

In order to  covariantly  formulate kinetic theory on NC spacetimes it is necessary to be able to describe the dynamics of pointlike particles propagating on such spacetime. To do so, we first notice that for a given curve $\gamma(\lambda)$, where $\lambda$ is the curve's parameter, and tangent vectors $\boldsymbol q=\dot\gamma(\lambda)$, the clock form $\tau=\tau_\mu dx^\mu$ naturally defines a proper time
\begin{equation}
    t(\gamma) = \int_\gamma \tau\,,\label{eq:Propper_time}
\end{equation}
which implies that  $\dot t=\tau_\mu \dot x^\mu(\lambda)$, where dot means derivative with respect to the parameter $\lambda$. Therefore, we define affine parameters as the ones for which
\begin{equation}
    \dot t = \text{const.} \implies \tau_\mu \dot x^\mu(\lambda) = \text{const.}\,.
\end{equation}
Thus, any affine parameter is related to the proper time as $\lambda=\alpha t + \beta$. We then define the reparametrization-, diffeomorphism-, and Milne-invariant action \cite{Kuchar:1980tw}
\begin{equation}
\begin{split}
   S&\equiv\int_\gamma d\lambda~\mathcal{L}\,,\\
   \mathcal{L}& = \frac{m}{2}\frac{h_{\mu\nu}}{\tau_\rho \dot x^\rho}  \dot x^\mu \dot x^\nu  +m m_\mu\dot x^\mu \,,
    \label{eq:non-rel-lagran1}
\end{split}
\end{equation}
where $m$ is the mass of the particle. 
Notice that under gauge transformations, the Lagrangian $\mathcal{L}$ transforms as a total derivative. 
The affinely parametrized ($\tau_\mu \dot x^\mu =1$) trajectories satisfy the (non-geodesic) equations of motion 
\begin{align}
    \dot x^\mu &= \frac{p^\mu}{m}\,,\\
 \dot p^\mu + \Gamma^{\mu}_{\alpha\beta}\,p^\alpha \dot x^\beta&=  - \frac{p^2}{2m} H^{\mu}\,_{\rho}\dot x^\rho \,,
    \label{eq:geodesic-with-N}
\end{align}
where $p^2 = h_{\mu\nu}p^\mu p^\nu$, and  $p^\mu$ is the kinematic momentum satisfying the constraint $\tau_\mu p^\mu=m$. Notice that the right-hand side of Eq. \eqref{eq:geodesic-with-N} resembles a Lorentz force on the particle after interpreting $-p^2/2m$ as the corresponding charge. The canonical momentum reads~\cite{Barducci:2017mse,Kluson:2017pzr,Banerjee:2019knz}
\begin{align}
    \pi_{\mu}= h_{\mu\nu}p^\nu + m m_\mu - \frac{p^2}{2m}\tau_\mu    \,.
    \label{eq:non-rel-momentum}
\end{align}
However, not all components of $\pi_\mu$ are independent\footnote{This can be traced to the reparametrization invariance of the action, which implies that the theory has a Hamiltonian constraint.}. In fact, in a $(d+1)$-dimensional spacetime, $\pi_\mu$ has only $d$ independent components due to the phase space constraint 
\begin{equation}
(\pi_\mu -m m_\mu) v^\mu + \frac{1}{2m}  (\pi_\nu -m m_\nu)(\pi^\mu -m m^\mu) = 0\,,
\end{equation}
as can be verified from Eq. \eqref{eq:non-rel-momentum}. In addition, the action is invariant under gauge transformations modulo a total derivative, which implies that the canonical momentum transforms as a gauge field
\begin{align}
    \pi_\mu \rightarrow \pi'_\mu=\pi_\mu + m\partial_\mu\Lambda\,.
\label{eq:eq:trans-of-moment}
\end{align}
Therefore, we find it more convenient to invert Eq. \eqref{eq:non-rel-momentum} (making use of $\tau_\mu p^\mu=m$) as
\begin{equation}
    p^\mu = \pi^\mu - m m^\mu + m v^\mu\,,
\end{equation}
and use the non-canonical pair $(x^\mu,p^\mu)$ as phase space coordinates, with $p^\mu$ obeying the constraint $\tau_\mu p^\mu = m$.

Thus, the time evolution of any phase space function $F(x,p)$ is computed as
\begin{equation}
    \dot F = X_L[F]\,,
\end{equation}
where $X_L$ is the Liouville operator defined as 
\begin{equation}
    mX_L =  p^\mu \frac{\partial}{\partial x^\mu} - \left( \Gamma^{\mu}_{\alpha\beta}p^\alpha p^\beta  + \frac{1}{2m}p^2H^{\mu}\,_{\rho}p^\rho\right)\frac{\partial}{\partial p^\mu}\,.
\label{eq:liouville}
\end{equation}
Notice that $X_L$ is tangent to the physical phase space, since $X_L[\tau_\mu p^\mu]=0$.

For generic geometries, the system will not have any Noether charge; however, our ultimate goal is to construct a kinetic theory, where a gas of interacting NC particles can equilibrate. In that case, it is mandatory to restrict the problem to geometries with at least one time-like Killing transformation. Therefore, we assume the existence of a set of parameters $\chi_K = \left(\xi_K^\mu \partial_\mu,\psi^K_{\mu}dx^\mu, \Lambda_K\right)$ such that the NC data is invariant under the action of $\delta_{\chi_K}$, i.e.,
\begin{equation}
    \delta_{\chi_K} \tau_{\mu}=\delta_{\chi_K} h^{\mu \nu} = \delta_{\chi_K} v^{\mu}=\delta_{\chi_K}m_{\mu}=0\,.
    \label{eq:killing_conds}
\end{equation}
Since the NC data are invariant under the transformation $\chi_K$, then it is clear that the action in Eq.~\eqref{eq:non-rel-lagran1} is also invariant under this variation, up to a possible boundary term, and there will exist a corresponding Noether charge \cite{banados_short_2016}.

Let us consider an arbitrary path deformation \( x^\mu(\lambda) \rightarrow x^\mu(\lambda) + \tilde{\xi}^\mu(x(\lambda)) \). The on-shell variation of the Lagrangian (see, e.g., \cite{banados_short_2016}) is expressed as
\begin{align}
    \delta_{\mathrm{on}} \mathcal{L} &= \tilde{\xi}^\mu  \frac{\partial \mathcal{L}}{\partial x^\mu}   +\dot{\tilde{\xi}}^\mu\frac{\partial \mathcal{L}}{\partial \dot x^\mu}=\frac{d}{d\lambda}\left(\tilde{\xi}^\mu\frac{\partial \mathcal{L}}{\partial \dot x^\mu}\right)\,,
    \label{eq:delta_L1}
\end{align}
In the last equality, we have used the Euler--Lagrange equations. On the other hand, the symmetry variation of the Lagrangian \( \mathcal{L} \) given in Eq.~\eqref{eq:non-rel-lagran1} under a Killing transformation \( \xi_K^\mu \) reads 
\begin{align}
    \delta_{\mathrm{sy}} \mathcal{L}&= \frac{m}{2\lb \tau_\rho \dot x^\rho\rb} \dot x^\alpha \dot x^\beta \pounds_{\xi_K} h_{\alpha\beta} -\frac{m}{2\lb \tau_\rho \dot x^\rho\rb^2}h_{\alpha\beta}\dot x^\alpha \dot x^\beta \dot x^\mu \pounds_{\xi_K} \tau_\mu + m \dot x^\alpha \pounds_{\xi_K} m_\alpha
    \nn
    &=-m\frac{d}{d\lambda}\Lambda_K\,,
    \label{eq:delta_L2}
\end{align}
The details of this procedure are discussed in \cite{vujanovic_group-variational_1970,djukic_procedure_1973,Sarlet_Generalizations_1981,manton_topological_2004}; for a detailed calculation, see Appendix~\ref{app:noether}. Now, if we replace \( \tilde{\xi}^\mu \) by \( \xi_K^\mu \) in the on-shell variation in Eq.~(\ref{eq:delta_L1}), then the left-hand sides of both variations are equal~\cite{banados_short_2016}. Comparing Eqs.~\eqref{eq:delta_L1} and \eqref{eq:delta_L2}, we obtain the conserved charge as 
\begin{align}
    Q_K=\xi_K^\mu \pi_\mu + m\Lambda_K\,,
    \label{eq:Q}
\end{align}
Note that \( Q_K \) is a scalar that is both Milne- and gauge-invariant.

In addition, we find it convenient to fix units such that Planck's, Boltzmann's, and the gravitational constants take values $h=k_B=G_N=1$. In these units, the relevant physical quantities of the problem can be expressed in powers of length only and read $[mass]=\mathrm{length}^{-1/3}$, $[momentum]=\mathrm{length}^{-1}$, and  $[energy]=[temperature]=[time]^{-1}=\mathrm{length}^{-5/3}$.

\section{Kinetic theory and conservation equations}
\label{sec:kinetic}

Kinetic theory models the statistical behavior of large systems of particles, typically in gases, by studying their microscopic motion. The central object in kinetic theory is the phase space distribution function $f(x,p)$
, which gives the number of particles $dN$  contained in a space-like volume at point $x$ and having momenta within a small range around $p$. Since the number of particles is an observable, the phase space volume needs to be invariant under all the coordinate and frame ambiguities of NC spaces. In other words, $dN=f\,\Omega_{2d}$ for some invariant volume $2d$-form $\Omega_{2d}$, which we will now construct. To begin with, notice that the bi-linear form $G_{\mu\nu}$ induces a volume form on the phase space
\begin{equation}
    \Omega_{2d+1} =   \omega_{d+1}\wedge \sigma_{d+1}\delta(\tau_\mu p^\mu - m)\,,
\end{equation}
where 
\begin{align}
    \sigma_{d+1} = \sqrt{G} dp^{01\ldots d} \,
\end{align}
is a volume element on the tangent space to the point $x$ and $\omega_{d+1}$ is the spacetime volume formed introduced in Eq.~\eqref{eq:VolNCspace}. This is analogous to the relativistic case \cite{frolov_introduction_2011,choquet-bruhat_introduction_2015,acuna-cardenas_introduction_2022}. Another important ingredient in kinetic theory is Liouville's theorem, which guarantees the conservation of the flux of trajectories crossing bounded space-like domains of the manifold. Therefore, it is necessary that $\Omega_{2d}$ satisfies $\pounds_{X_L}\Omega_{2d}=0$ and $d\Omega_{2d}=0$. In fact, the following form satisfies such properties \cite{stewart1971non}
\begin{align}
    \Omega_{2d} &= X_{L}\cdot\Omega_{2d+1}
\,.
\end{align}
Explicitly, $\Omega_{2d}$ is
\begin{equation}
    \Omega_{2d} = \frac{p^\mu}{m}\Sigma_\mu\wedge \sigma_{d+1}\delta(\tau_\mu p^\mu - m)+\Omega'_{2d}\,,
\end{equation}
where $\Sigma_\mu=\frac{1}{d!}\epsilon_{\mu\mu_1\ldots\mu_d} dx^{\mu_1\ldots\mu_d}$ is the $d-$dimensional surface element and $\Omega'_{2d}$ is a form proportional to $\omega_{d+1}$. Thus,  $\Omega'_{2d}$ vanishes when acting on space-like vector fields, and as a consequence $\Omega'_{2d}$ will not play a role in the kinetic theory integrals we will consider, as we shall see later.

A key ingredient of kinetic theory is the Boltzmann equation, which describes how the distribution function $f(x,p)$ evolves under the dynamics. More precisely, the Boltzmann equation is
\begin{align}\label{eq:Boltzmann}
    X_L[f]=\mathcal{C}[f]\,,
\end{align}
where the left-hand side accounts for the non-interacting dynamics of the particles and is determined by the single-particle equations of motion, while $\mathcal{C}[\cdot]$ is called the collision operator and it contains information about the interactions between the microscopic constituents. We require $\mathcal{C}[\cdot]$ to obey the following properties:
\begin{itemize}
    \item It is invariant under the NC transformations. 
    \item Only scattering processes involving two particles contribute to the collision integral.
    \item The number of particles before and after a collision does not change.
    \item Interactions are instantaneous and local.
    \item The total energy and momentum in a collision are conserved.\footnote{We assume that the geometry fields vary slowly compared to the distances and time intervals at which the interactions happen.}
\end{itemize}

\subsection{Currents and conservation laws}

The total mass in a space-like volume $V$ of the spacetime is
\begin{align}
    M &= m\int_{V\times T_V} f\,\Omega_{2d}=m\int_V \mathcal{J}^\mu (x)\Sigma_\mu\,,
\end{align}
where the particle current is defined as
\begin{align}
    \mathcal{J}^\mu(x) &= \int_{T_x} \frac{p^\mu}{m} f\,\sigma_{d+1}\delta(\tau_\mu p^\mu - m)\,,\label{eq:Jmu}
\end{align}
with $T_x$ the tangent space at the point $x$. Associated to the observer with velocity $v^\mu$, we can define the amount of kinematic momentum $P_\mu$ and energy contained in a volume $V$
\begin{align}
    P_\mu &= \int_{V\times T_V} p_\mu f\,\Omega_{2d} = \int_V \mathcal (\mathcal P_\mu v^{\nu} + T_{\mu\alpha}h^{\alpha\nu})\Sigma_\nu \equiv \int_V \mathcal T_\mu\,^{\nu} (x)\Sigma_\nu\,,\\
     E &= \frac{1}{2m}\int_{V\times T_V} p^2\,f\,\Omega_{2d}= \int_V\tilde{\mathcal{E}}^\mu(x) \Sigma_\mu \equiv  \int_V\mathcal{E}^\mu(x) \Sigma_\mu+ \int_V\left(u_\alpha - \frac{1}{2}u^2\tau_\alpha\right)\mathcal T^{\mu\alpha}\Sigma_\mu     \,, \label{eq:E}
\end{align}
where the corresponding currents are defined as
\begin{align}
    \mathcal{T}^{\mu\nu}(x) &=  \int_{T_x}  \frac{p^\mu p^\nu}{m}f\, \sigma_{d+1}\delta(\tau_\mu p^\mu - m)\,,\label{eq:Tmunu}\\
    \mathcal{E}^{\alpha}(x) &= \frac{1}{2m}g_{\mu\nu}\int_{T_x}  \frac{p^\mu p^\nu p^\alpha}{m}f\, \sigma_{d+1}\delta(\tau_\mu p^\mu - m)\,.\label{eq:Emu}
\end{align}
When writing Eqs.~(\ref{eq:E}) and~(\ref{eq:Emu}), we exploited the fact that for matter at non-zero density there always exists a velocity field $u^\mu$ associated to the rest frame, so that the spacetime has the structure of a TNC manifold described in Sec.~\ref{sec:TNC}. Consequently, the energy current $\mathcal E^\mu$ has been defined so that the observer-ambiguous contribution has been subtracted. The energy current $\mathcal E^\mu$ is Milne invariant and can be interpreted as the energy flux in the fluid's co-moving frame. Moreover, we have decomposed the invariant ``thermal" stress-energy tensor $\mathcal T^{\mu\nu}$ into the frame-dependent momentum density $\mathcal P_\mu $ and the stress tensor $T_{\mu\nu}$. The invariant and frame-dependent currents are related as follows:
\begin{align}
    \mathcal T^{\mu\nu} &= h^{\mu\alpha}h^{\nu\beta}T_{\alpha\beta} + h^{\mu\alpha}\mathcal P_\alpha v^\nu + h^{\nu\alpha}\mathcal P_\alpha v^\mu 
    + m(\tau_\rho \mathcal J^\rho)\, v^\mu v^\nu\, ,\\
   \tilde{ \mathcal E}^\mu &= \mathcal{E}^\mu + \mathcal T^{\mu\alpha}\left(u_\alpha - \frac{1}{2}u^2\tau_\alpha\right)\,.
\end{align}
As we show below, in the hydrodynamic regime, all the microscopic information of the system is encoded in constitutive relations for the invariant currents. 
Also notice that the particle current  and momentum density obey the Milne Ward identity
\begin{equation}
    \mathcal P_\mu  = m h_{\mu\nu}\mathcal J^\nu\,.
\end{equation}

Typically, in kinetic theory it is useful to introduce the higher moments of the distribution function, which in our NC description can be defined as 
\begin{align}
    \mathcal{I}^{\alpha_1\alpha_2\cdots\alpha_n}=\frac{1}{m^n}\int f p^{\alpha_1} p^{\alpha_2} \cdots p^{\alpha_n} \sigma_{d+1}\delta(\tau_\mu p^\mu - m)\,,
    \label{eq:moments}
\end{align}
which satisfy the equations (see Appendix~\ref{app:deriv_volume} for a detailed derivation)
\begin{equation}
\begin{split}
    \left(\bar\nabla_{\mu}+H_{\mu \nu}u^\nu\right)\mathcal{I}^{\mu\alpha_2\cdots\alpha_n} = & \frac{1}{m^{n-1}}\int p^{\alpha_2} \cdots p^{\alpha_n} X_L[f]\sigma_{d+1}\delta(\tau_\mu p^\mu - m)\\
    &-\frac{n-1}{2}H^{(\alpha_2}_{~~~\rho}g_{\beta\gamma}\mathcal{I}^{\beta\gamma\rho|\alpha_3\cdots\alpha_n)}\,. 
    \end{split}
    \label{eq:moments_covariant_0}
\end{equation}
If $f$ is a solution of the Boltzmann equation $X_L[f]=\mathcal{C}[f]$, Eq.~(\ref{eq:moments_covariant_0}) becomes
\begin{equation}
    \left(\bar\nabla_{\mu}+H_{\mu \nu}u^\nu\right)\mathcal{I}^{\mu\alpha_2\cdots\alpha_n} = \mathcal{K}^{\alpha_2\cdots\alpha_n}-\frac{n-1}{2}H^{(\alpha_2}\,_{\rho}g_{\beta\gamma}\mathcal{I}^{\beta\gamma\rho|\alpha_3\cdots\alpha_n)}\,, 
    \label{eq:moments_covariant}
\end{equation}
where 
\begin{equation}
    \mathcal{K}^{\alpha_1\cdots\alpha_n} = \frac{1}{m^{n}}\int p^{\alpha_1} \cdots p^{\alpha_n} \mathcal{C}[f]\sigma_{d+1}\delta(\tau_\mu p^\mu - m)\,.
    \label{eq:K_define}
\end{equation}

We can identify
\begin{equation}
    \mathcal{J}^\mu = \mathcal{I}^\mu\,,~~~~\mathcal{T}^{\mu\nu} = m\mathcal{I}^{\mu\nu}\,,~~~~\mathcal{E}^\alpha = \frac{m}{2}g_{\mu\nu}\mathcal{I}^{\mu\nu\alpha}\,.
    \label{eq:currents_define}
\end{equation}
The requirement that collisions conserve particle number, momentum, and energy, imply the following properties of the collision operator:
\begin{equation}
    \mathcal{K}=0\,,~~~~\mathcal{K}^\mu=0\,,~~~~g_{\mu\nu}\mathcal{K}^{\mu\nu}=0\,.
    \label{eq:c_conditions}
\end{equation}
Thus, the equations of motion for the mass, momentum and energy currents reduce to the set of conservation equations (with the derivation of the equation for $\mathcal{E}^\mu$ relegated to Appendix \ref{app:energy}):
\begin{align} 
    \left(\bar\nabla_\mu+H_{\mu\rho}u^\rho\right)\mathcal{J}^{\mu}&=0\,, 
    \label{eq:Wards_j}
    \\
    \left(\bar\nabla_\mu+H_{\mu\rho}u^\rho\right)\mathcal{T}^{\mu\nu}&=-H^\nu_{~~\rho}\mathcal{E}^\rho\,, 
    \label{eq:Wards_t}
    \\
    \left(\bar\nabla_\mu+ H_{\mu\rho}u^\rho\right)\mathcal{E}^\mu&= -H_{\mu\rho}u^\rho\mathcal{E}^\mu -\mathcal{T}^{\mu\nu}g_{\rho(\mu}\bar\nabla_{\nu)}u^\rho \,. \label{eq:Wards_e}
\end{align} 
We see that \eqref{eq:Wards_j} and \eqref{eq:Wards_t} are not independent since the former is obtained after contracting the latter with the clock form $\tau_\mu$. In addition, the momentum and energy conservation equations \eqref{eq:Wards_t}, \eqref{eq:Wards_e}, show energy-related ``Lorentz force" and ``Joule heating", respectively, consistent with the single-particle equations of motion \eqref{eq:geodesic-with-N}. Most importantly, Eqs. \eqref{eq:Wards_j}, \eqref{eq:Wards_t}, and \eqref{eq:Wards_e} agree with the Ward identities derived in Refs. \cite{Jensen:2014aia, Jensen:2014ama} using a field-theoretical approach, showing the reliability of our kinetic theory. We emphasize Eqs. \eqref{eq:Wards_j}, \eqref{eq:Wards_t} and \eqref{eq:Wards_e} contain the same information as the set of equations that we could have obtained if we had worked with the frame-dependent currents.

\subsection{Detailed balance and thermal equilibrium }
\label{sec:equilibrium}

Before introducing the notion of equilibration, we find it convenient to define the entropy
\begin{equation}
    S = -\int_{V\times T_V} f\log f\,\Omega_{2d}\,,
    \label{eq:Smu}
\end{equation}
where $V$ is a space-like volume of the spacetime and $T_V$ is the tangent space to $V$. The entropy satisfies  
\begin{equation}
    \dot{S} = X_L[S] \geq 0\,.
    \label{eq:EntropyProduction}
\end{equation}
We define a local equilibrium state by a distribution function $f_0$ that saturates the inequality in Eq. \eqref{eq:EntropyProduction}, i.e., $\dot S = 0$. This happens if $\log f_0$ is a collisional invariant, so that $f_0$ obeys the detailed balance condition $\mathcal{C}[f_0]=0$. We will call $f_0$ a hydrodynamic distribution function. According to the above discussion,
 \begin{equation}
    f_{0}(x,p) \equiv \mathcal N\exp\left[Q_{0}(x,p)\right]\,,
    \label{eq:local_def_1}
\end{equation}
with $\mathcal{N}$ a normalization constant and  $Q_{0}(x,p)$ the collisional invariant
\begin{equation}
Q_{0} = \alpha(x) + \xi^\mu(x)h_{\mu\nu}p^\nu - \gamma(x) p^2\,,
\end{equation}
where $\alpha(x)$, $\gamma(x)$, and $\xi^\mu(x)$ are arbitrary functions constrained by the requirement that $Q_0$ must be a scalar. By completing the square, $Q_0$ can be expressed as  
\begin{equation}
    Q_{0} =-\frac{1}{2mT} (p-m u)^2 +\frac{\mu}{T}\,,
    \label{eq:dist_collision}
\end{equation}
where $2mT=\gamma^{-1}$, $u^\mu=T\xi^\mu$, and $\mu=T\alpha+mu^2/2$. The invariance of $Q_0$ under Milne boost fixes $u^\mu\tau_\mu=1$\footnote{Under Milne boost $\delta Q_0 \propto  (p^\mu-mu^\mu)\tau_\mu$, which has to vanish on the mass shell.}, which implies the useful relations
\begin{equation}
    T = \frac{1}{\tau_\mu \xi^\mu}\,,~~~~u^\mu = T \xi^\mu\,.
    \label{eq:T-and-u}
\end{equation}
Finally, the collisional invariant can be expressed in the form of Eq. \eqref{eq:Q},
\begin{align}
    Q_{0}=\xi^\mu \pi_\mu + m\Lambda\,,~
    \label{eq:Q_general}
\end{align}
with the gauge parameter $\Lambda$ satisfying the relation
\begin{equation}
    \frac{\mu}{mT} = \xi^\mu A_{\mu} + \Lambda\,.
    \label{eq:mu_inv}
\end{equation}
We interpret the parameters $T,u^\mu,\mu$ as hydrodynamic temperature, fluid velocity and chemical potential, respectively.

On the other hand, a generic hydrodynamic distribution function $f_0$ will not be a solution of the full Boltzmann equation. In addition to requiring $\mathcal{C}[f_{\mathrm{eq}}]=0$, the equilibrium distribution $f_{\mathrm{eq}}$ has to be a constant of motion satisfying
\begin{equation}
    X_L[f_{\mathrm{eq}}]=0\,.
\end{equation}
Following an explicit calculation of $X_L[f_0]$ in Appendix \ref{app:deriv_LX_Q}, we find that
\begin{equation}
   mX_L[f_0] = f_0\lb x,p\rb\left(m p^\beta \delta_{\chi} A_\beta-\frac{1}{2}p^\alpha p^\beta g_{\alpha\nu}g_{\beta\mu}\delta_\chi h^{\mu\nu} -\frac{1}{2m} h_{\alpha\beta}  p^\alpha p^\beta p^\gamma \delta_{\chi}\tau_\gamma\right)\,,
   \label{eq:LX_f}
\end{equation}
where the relation between $\chi=(\xi^\mu\partial_\mu,0,\Lambda)$ and the thermodynamic variables $(T,u^\mu,\mu)$ is fixed by Eqs.~(\ref{eq:T-and-u}) and (\ref{eq:mu_inv}). If $X_L[f_0]=0$ is to be valid for all momenta, we need to have 
\begin{equation}
   \delta_\chi \tau_\gamma =\delta_\chi h^{\alpha\beta}=\delta_\chi A_{\mu}=0\,,
   \label{eq:killing_invariant}
\end{equation}
which is equivalent to requiring that $\chi=\chi_K$ with $\chi_K$ being a Killing transformation, see Eq.~(\ref{eq:killing_conds}). Thus, $f_{\mathrm{eq}}=\exp\left[Q_K(x,p)\right]$. In fact, the Killing conditions in Eq.~(\ref{eq:killing_invariant}) can be equivalently expressed in terms of the local thermodynamic variables $(T,u^\mu,\mu)$ as hydrostatic conditions, derived in Appendix~\ref{app:killing_conditions}:
\begin{align}
  & u^\mu\partial_\mu  T =0  \,, \qquad (\partial^\mu   +  H^\mu\,_{\nu} u^\nu )T=0\,, \label{eq:killing_tau}\\
  & u^\mu\partial_\mu \mu =0 \,, \qquad  T\partial^\mu\left(\frac{\mu}{T}\right)  + m a^\mu=0\,, \label{eq:mu_hydro}\\
  & \theta =0\,, \quad\quad\qquad \sigma^{\mu\nu} =0\,. \label{eq:sigma-theta}
\end{align}
Note that temperature, fluid velocity, and chemical potential are well defined variables only at equilibrium where they are entirely fixed by the isometries of the space-time, and that on manifolds without time-like Killing symmetry equilibration is not possible.

In the next section, we show how the formalism introduced in this section can be used to find the equilibrium distribution function of a gas in a non-inertial frame of reference. Then, in Sec.~\ref{sec:hydro} we will start exploring out-of-equilibrium physics by deriving the equations of Milne-invariant hydrodynamics.

\section{Rotating gases: rest frame and equilibrium}
\label{sec:rotate}

\begin{figure*}
\includegraphics[width=0.7\textwidth]{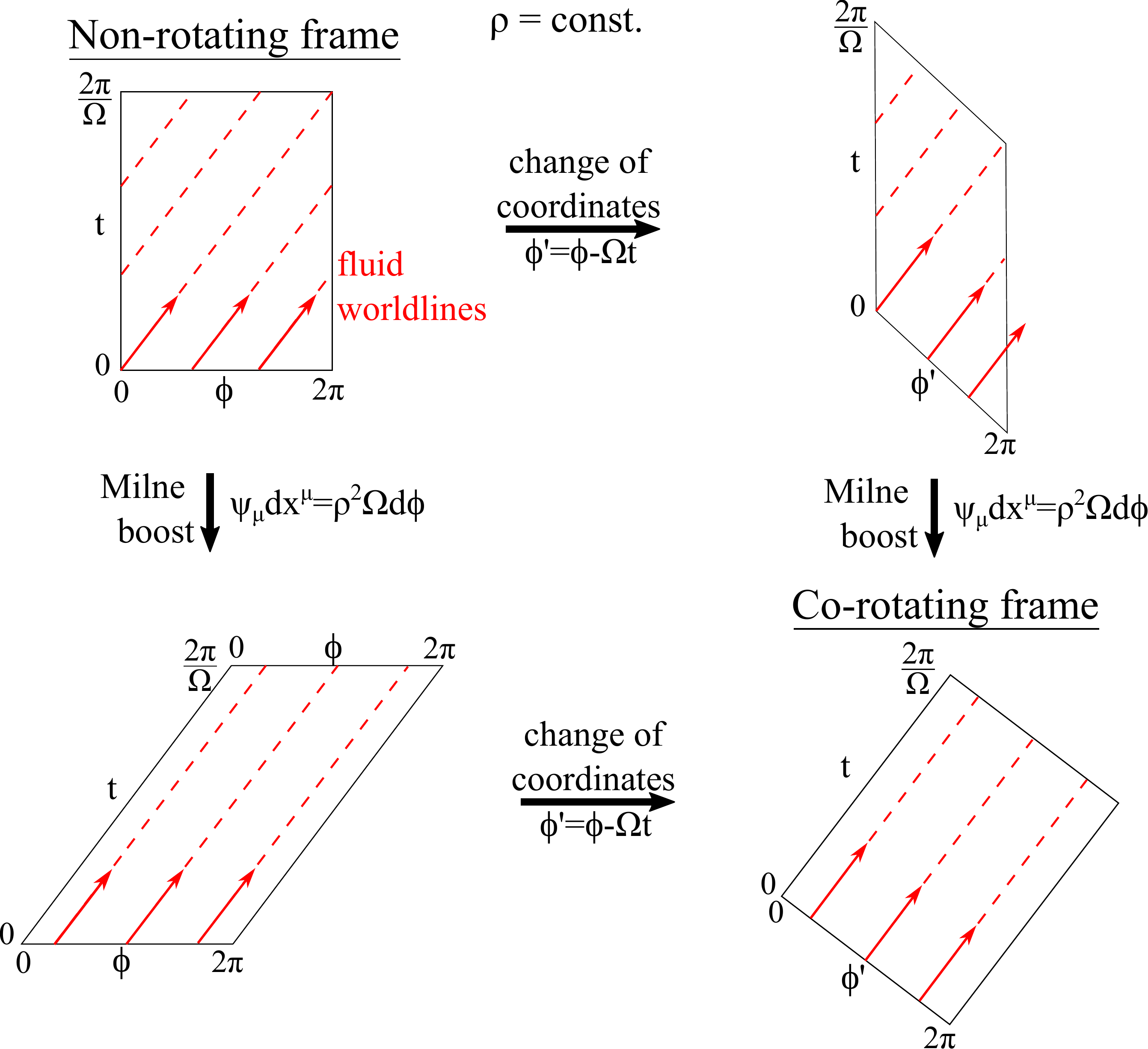}
\caption{Shifting to the reference frame co-rotating with the fluid, as described in Sec.~\ref{sec:rotate}. The worldlines of the fluid elements (red dashed lines) at a fixed $\rho$ are plotted for different coordinate systems: $(t,\phi)$ and $(t,\phi')$, and different choices of the reference velocity: $v^\mu$ and $v'^\mu$. The tilt of the $t$ axis is given by the angle between $\partial_t$ and $\partial_\phi$ (left column) or $\partial_{\phi'}$ (right column) calculated using the metric $h_{\mu\nu}$ (top row) or $h'_{\mu\nu}$ (bottom row).}
\label{fig:plot}
\end{figure*}

The primary goal of this section is to highlight the crucial role played by the Milne boost in shifting between the rest frames of different observers. For simplicity, we will consider a two-dimensional NC spacetime\footnote{The discussion straightforwardly generalizes to the three-dimensional case.} of the form \eqref{eq:flatNC} deformed by a static rotationally invariant scalar potential. In addition, we will assume a gas rotating rigidly on such a geometry with angular velocity $\Omega$. In polar coordinates $(t,\rho,\phi)$, the fluid velocity reads
\begin{equation}
    u^\mu\partial_\mu = \partial_t + \Omega\partial_\phi\,.
\end{equation}
The Milne invariant fields for such spacetime are $\tau=dt$ and $h^{\mu\nu}\partial_\mu\otimes\partial_\nu=\partial_\rho\otimes\partial_\rho + \rho^{-2}\partial_\phi\otimes\partial_\phi$. In the first frame to be considered, the Milne dependent fields read
\begin{equation}
    h_{\mu\nu}dx^\mu dx^\nu = 
   d\rho^2 +\rho^{2}d\phi^2, ~~~~v^\mu\partial_\mu = 
    \partial_t, ~~~~m_\mu dx^\mu = 
        -\varphi(\rho)dt\,.
\end{equation}
Notice that relations \eqref{eq:acceleration_curl} imply that the vorticity of such family of observers vanishes, whereas their geodesic accelaration is $a_v^\mu=\varphi'\delta^\mu_1$.

On the other hand, we notice that by using the Milne parameter $\psi_\mu dx^\mu= \rho^2\Omega d\phi$, we can change the frame such that the new reference velocity is $v'^\mu=u^\mu=v^\mu + h^{\mu\nu}\psi_\nu$. By construction, on such a frame the geodesic accelaration $a_{v'}^\mu$ and vorticity $\omega_{v'}^{\mu\nu}$ will coincide with the ones of the fluid. In the co-moving frame, the primed Newton-Cartan data reads
\begin{equation}
    h'_{\mu\nu}dx^\mu dx^\nu = 
   d\rho^2 +\rho^{2}(d\phi-\Omega dt)^2, ~~~~v'^{\mu}\partial_\mu = 
    u^{\mu}\partial_\mu, ~~~~m'_\mu dx^\mu = 
   -    \left(\varphi(\rho) - \frac{1}{2}\rho^2\Omega^2  \right)dt + \rho^2\Omega\left(d\phi-\Omega dt\right)\,.
\end{equation}
However, the worldlines of the fluid elements are not given by paths with a constant $\phi$ since $u^\mu\partial_\mu\phi\neq 0$, see Fig.~\ref{fig:plot}. To define the rest frame of the fluid, we require that in addition to adjusting the reference velocity such that $v^\mu=u^\mu$, the coordinate system has to be also adjusted such that $u^\mu\partial_\mu x^\nu = \delta^\nu_0$. In our problem, after performing the change of coordinates $t'=t$, $\rho'=\rho$, and $\phi'=\phi-\Omega t$, the basis $\partial_{\rho'},\partial_{\phi'}$ gets locked to the motion of the gas, and the NC fields become
\begin{equation}
    h'_{\mu\nu}dx'^\mu dx'^\nu = 
   d\rho'^2 +\rho'^{2}d\phi'^2, ~~~~v'^{\mu}\partial'_\mu = 
    \partial_{t'}, ~~~~m'_\mu dx'^\mu = 
   -    \left(\varphi(\rho) - \frac{1}{2}\rho^2\Omega^2  \right)dt' + \rho^2\Omega d\phi'\,.
\end{equation}
After calculating the field strength $F'_{\mu\nu}=\partial_\mu m'_\nu-\partial_\nu m'_\mu$, the geodesic accelaration and vorticity can be found as
\begin{align}
    a_{v'}^\mu &=  (\varphi' -\rho'\Omega^2)\delta_1^\mu\,,\\
    \omega_{v'}^{\mu\nu} &= 2\Omega\,u^\alpha\epsilon_\alpha\,^{\mu\nu}\,.
    \label{eq:vorticity_rotating}
\end{align}
Remarkably, this geometric formulation naturally encodes the similarity of the non-inertial effects (the so-called centrifugal and Coriolis forces) to the Lorentz force of electrodynamics. Notice also that if the scalar potential is harmonic such that $\varphi=1/2\Omega\rho^2$, the centrifugal force will cancel with the central force and the accelaration vanishes. 

This example illustrates how the presence of the gas naturally allows us to extend the original NC structure into the thermal NC spacetime introduced in Sec. \ref{sec:TNC} using the fluid velocity. In fact, the Milne invariant fields $g_{\mu\nu},A_\mu$ correspond to $h'_{\mu\nu},m'_\mu$, respectively, in our construction. Therefore, in the thermal NC language, the fluid rest frame is given by the geometric data $(\tau_\mu,u^\mu,h^{\mu\nu},g_{\mu\nu},A_\mu)$ expressed in coordinates such that $u^\mu=\delta^\mu_0$.

In addition, since the spacetime considered here possesses Killing vectors $\partial_t,\partial_\phi$, and the fluid velocity is a linear combination of them, $u^\mu$ is a Killing vector. A straightforward calculation reveals that the hydrostatic conditions in Eqs. \eqref{eq:killing_tau}-\eqref{eq:sigma-theta} are satisfied if the chemical potential and temperature are
\begin{align}
T & = \text{const}\,,\\
\mu &
=-    \varphi(\rho) + \frac{1}{2}\rho^2\Omega^2 +\text{const}  \,. \label{eq:mu_rotating}
\end{align}
This relation implies that in flat space, finite particle density at infinity requires the scalar potential to diverge as $1/2\Omega\rho^2$, with the fine-tuned case $\varphi = 1/2\Omega\rho^2$ corresponding to an equilibrium state with uniform density. Contrary to relativistic gases where causality forbids equilibrium states with rigid rotation, non-relativistic gases have no such formal obstruction, although such equilibrium can be considered unphysical due to the divergence of the particles' density at the spatial infinity.

\section{NC hydrodynamics from kinetic theory }
\label{sec:hydro}
One of the most important contributions of kinetic theory is that it provides a microscopic foundation for hydrodynamics of weakly coupled gases. Therefore, in order to lay the groundwork for future developments and uses of our formalism, in this section we derive equations of Newton-Cartan hydrodynamics and fix the first-order transport coefficients in two- and three-dimensional gases. The results are compared to analogous non-covariant results scattered in the literature.

\subsection{Zeroth-order hydrodynamics}\label{sec:idealhydro}

From this section onwards, we specify to the torsionless case $H_{\mu\nu}=0$ for simplicity. Importantly, this also removes the distinction between the different covariant derivatives, so that $\bar{\nabla}_\mu = \nabla_\mu$.

Let us start our discussion of hydrodynamics by calculating the ideal particle current $\mathcal{J}^\mu_{0}$, stress tensor $\mathcal{T}^{\mu\nu}_{0}$, and energy current $\mathcal{E}^{\mu}_{0}$, defined with respect to the hydrodynamic distribution function $f_0$ given by Eqs.~(\ref{eq:local_def_1}) and~(\ref{eq:dist_collision}). Evaluating the Gaussian integral gives
\begin{equation}
    \mathcal{J}^\mu_{0} = \sqrt{G}m^{-1}\int d^{d+1} p\,\delta\lb\tau_\rho  p^\rho-m\rb p^\mu f_{0} = c\,\mathcal{N} \lb 2\pi m T\rb^{d/2}e^{\mu/T}u^\mu\equiv n u^\mu\,.
    \label{eq:current_ideal}
\end{equation}
Similarly, the other currents can be evaluated as
\begin{equation}
\begin{split}
\mathcal{T}^{\mu\nu}_{0} & = \rho \,u^\mu u^\nu + \mathfrak p h^{\mu\nu} \,,  \\
    \mathcal{E}^\mu_{0} & = \epsilon u^\mu\,,
\end{split}    \label{eq:currents_id}
\end{equation}
where the mass density, internal energy and pressure are $\rho=mn$, $\epsilon= \frac{dnT}{2}$ and  $\mathfrak p=n T$ respectively. On the other hand, the ideal constitutive relations for an observer with velocity $v^\mu$  are
\begin{align}
    \mathcal P_\mu &= \rho \,u_\mu \,,\qquad T_{\mu\nu} = \mathfrak p\, h_{\mu\nu}  \,,\\
    \tilde{\mathcal E}^\mu &= \left(\epsilon + \frac{\rho}{2}u^2 \right)u^\mu + \mathfrak p \,P^\mu_\nu\,u^\nu\,.
\end{align}
More generally, all the moments $\mathcal{I}_{0}^{\alpha_1\alpha_2\cdots\alpha_k}$ of $f_{0}$ can be calculated explicitly thanks to the Gaussian form of the distribution function. These moments consist of all fully symmetrized combinations of $u^\mu$ and $m^{-1}Th^{\mu\nu}$ multiplied by $n$.

As previously argued, $f_0$ solves the Boltzmann equation only in the hydrostatic regime. Therefore, Eqs. \eqref{eq:currents_id} are exact only at equilibrium. However, since hydrodynamics is based on the idea of local equilibrium, it is natural to expand the out-equilibrium  distribution function around the hydrodynamic distribution $f_{0}(x,p)$ evaluated at non-hydrostatic velocity, temperature and chemical potential defined as in \eqref{eq:T-and-u} and \eqref{eq:mu_inv}, so that $f(x,p)=f_{0}(x,p)+\delta f(x,p)$. Consequently, we split the moments of the distribution function $f(x,p)$ into $\mathcal{I}^{\alpha_1\alpha_2\cdots\alpha_k}=\mathcal{I}^{\alpha_1\alpha_2\cdots\alpha_k}_{0}+ \delta \mathcal{I}^{\alpha_1\alpha_2\cdots\alpha_k}$.
   In addition, we restrict the analysis to systems where the relaxation time approximation is applicable 
\begin{equation}
    \mathcal{C}[f] = -\frac{\delta f}{\tau_{\mathrm{coll}}}\,,
\end{equation}
with $\tau_{\mathrm{coll}}$ a phenomenological parameter that by construction is a scalar. With this choice we have
\begin{equation}
    \mathcal{K}^{\alpha_1\cdots\alpha_k} = -\frac{1}{\tau_{\mathrm{coll}}}\delta \mathcal{I}^{\alpha_1\alpha_2\cdots\alpha_k}\,.
    \label{eq:tra}
\end{equation}
However, (covariant) conservation of particle number, momentum and energy  requires $\delta \mathcal{I}=\delta \mathcal{I}^{\mu}=g_{\mu\nu}\delta \mathcal{I}^{\mu\nu}=0$, conditions that are automatically satisfy if we fix the Eckart frame~\cite{Eckart:1940te,Jensen:2014ama}. The ambiguity in the definition of the hydrodynamic variables ($\mu,T,u^\mu$) is fixed by requiring that the equations of state and the particle current do not receive higher derivative corrections
\begin{equation}
    n = c\mathcal N\lb 2\pi m T\rb^{d/2}e^{\mu/T}\,,\qquad \epsilon =\frac{d}{2} nT\,, \qquad\mathcal J^\mu = nu^\mu\,.\label{eq:eckart}
\end{equation}
Notice that in the discussion of hydrodynamic frames, the emphasis is usually placed on fixing the velocity. However, we stress that it is essential to fix all thermodynamic functions that are not uniquely defined out of equilibrium.

The out-of-equilibrium currents then can be expressed  as
\begin{align}
    \mathcal{T}^{\mu\nu} = &   \rho u^\mu u^\nu + \mathfrak p\, h^{\mu\nu} + \tau^{\mu\nu}\,,  \\
    \mathcal{E}^\mu = & \epsilon \, u^\mu + q^\mu\,,
\end{align}
with the higher order corrections to the currents a transverse tensor a vector respectively, defined as $\tau^{\mu\nu}\equiv m \delta \mathcal{I}^{\mu\nu}$ 
  $q^\mu \equiv \frac{m}{2} g_{\nu\lambda}\delta\mathcal{I}^{\mu\nu\lambda}$. In addition, the correction to the stress-momentum tensor is traceless $\tau^{\mu\nu}g_{\mu\nu}=0$.
  With these definitions, the conservation equations  (\ref{eq:Wards_j}), (\ref{eq:Wards_t}), and (\ref{eq:Wards_e}), can be rewritten as equations of motion for $\mu$, $u^\nu$, and $T$:
\begin{align}
    &u^\mu\partial_\mu n+n\nabla_\mu  u^\mu=0\,,\label{eq:eoms_thermo_covariant-a}\\
    &u^\mu\nabla_\mu u^\nu + h^{\mu\nu}\partial_\mu \mathfrak p +\nabla_\mu \tau^{\mu\nu}=0\,,\label{eq:eoms_thermo_covariant-b}\\
    & \frac{d}{2}nu^\mu\partial_\mu T+ nT\nabla_\mu u^\mu+\tau^{\mu\nu}g_{\rho\mu}\nabla_\nu u^\rho+\nabla_\mu q^\mu =0 \,.\label{eq:eoms_thermo_covariant-c}
\end{align}

Furthermore, the general equation of motion for the moments is given in Eq. (\ref{eq:moments_covariant}) can be rewritten as
\begin{equation}
\nabla_\mu\delta\mathcal{I}^{\mu\alpha_1\ldots\alpha_k}-\mathcal{K}^{\alpha_1\ldots\alpha_k}=\mathcal{S}^{\alpha_1\ldots\alpha_k}\,,
    \label{eq:moments_covariant_2}
\end{equation}
where we define 
\begin{equation}
    \mathcal{S}^{\alpha_1\ldots\alpha_k}\equiv -\nabla_\mu\mathcal{I}^{\mu\alpha_1\ldots\alpha_k}_{0}
\end{equation}
and from Eq. (\ref{eq:moments_covariant_0}), remembering that $H_{\mu\nu}=0$, we have 
\begin{equation}
    \mathcal{S}^{\alpha_1\ldots\alpha_k}= -\frac{1}{m^k}\int \sigma_{d+1}\,\delta\lb\tau_\nu  p^\nu -m\rb p^{\alpha_1}  \cdots p^{\alpha_k} X_L[f_0]\,.
    \label{eq:F_covariant}
\end{equation}
The term $X_L[f_{0}]$ was already calculated in Eq. (\ref{eq:LX_f}), where it was found to be proportional to $\delta_{\chi}h^{\mu\nu}$, $\delta_\chi A_\mu$ and $\delta_{\chi}\tau_\mu$. This explicitly proves that $\mathcal{S}^{\alpha\beta\ldots\gamma}$ vanishes when in equilibrium. 

Equations~\eqref{eq:eoms_thermo_covariant-a}-\eqref{eq:eoms_thermo_covariant-c} contain two unknown tensors, $\tau^{\mu\nu}$ and $q^\mu$, which have to be determined in order for the system of equations~~\eqref{eq:eoms_thermo_covariant-a}-\eqref{eq:eoms_thermo_covariant-c} to be solvable beyond the ideal order. In the hydrodynamic approach, this is done by expanding $\tau^{\mu\nu}$ and $q^\mu$ in spatial gradients of the hydrodynamic variables $T$, $\mu$ and $u^\mu$. A gradient expansion that generalizes the Chapman-Enskog method to a NC gas can be performed if the timescales of out-of-equilibrium perturbations are much larger than the timescales at which non-conserved quantities relax, so that local equilibration can be assumed. Due to the rather technical character of these calculations, we relegate the details of the gradient expansion procedure, as well as a more thorough discussion of the assumptions involved, to Appendix \ref{sec:gradient}.

\subsection{First-order constitutive relations}
\label{sec:gases}

In this section we present the constitutive relations for $\tau^{\mu\nu}$ and $q^\mu$ at the first order of the gradient expansion. Even though the transport coefficients calculated this way agree with the ones found in the literature, our formalism allows one to unambiguously determine the form of constitutive equations and the values of transport coefficients for rotating gases in a covariant manner. All the formulas in this section are derived in Appendix \ref{sec:gradient}.

\subsubsection{Two-dimensional rotating gas}

Let us first consider the hydrodynamic flow of a two-dimensional gas around an equilibrium state with non-vanishing vorticity. Therefore,
we define the \textit{vorticity pseudoscalar} 
\begin{equation}
    B = \frac{1}{2c}u^\rho \epsilon_{\rho\mu\nu}f^{\mu\nu}\,.
    \label{eq:vorticity_scalar}
\end{equation}
The vorticity scalar field, defined in this way, contains a contribution of the Coriolis force as measured in the rest frame of the fluid. In that case, we can identify $B=2\Omega$ with $\Omega$ the angular frequency, according to Eq.~(\ref{eq:vorticity_rotating}).
The first-order stress tensor can be written as
\begin{equation}
    \tau^{\mu\nu} = -\eta^{\mu\nu}_{\alpha\beta} \sigma^{\alpha\beta}\,.
    \label{eq:tau_2d}
\end{equation}
Since in two dimensions there exist two rotationally-invariant traceless symmetric tensors \cite{avron1998odd}, the viscosity tensor $\eta^{\mu\nu}_{\alpha\beta}$ can be written as a sum of two terms
\begin{equation}
    \eta^{\mu\nu}_{\alpha\beta} = \lambda^{(e)} \eta^{(e)\mu\nu}_{\alpha\beta}+\lambda^{(o)} \eta^{(o)\mu\nu}_{\alpha\beta}\,.
\end{equation}
The superscript $``(e)"$ corresponds to a parity-even viscosity tensor, while $``(o)"$ corresponds to a parity-odd one. Explicitly, 
\begin{equation}
    \eta^{(e)\mu\nu}_{\alpha\beta} = \bar P^{\langle \mu}_{\alpha}\bar P^{\nu\rangle}_{\beta}, ~~~~~~
    \eta^{(o)\mu\nu}_{\alpha\beta} = c^{-1}u^\rho \epsilon_{\rho\lambda\alpha}\bar P^{\langle\mu}_{\beta}h^{\nu\rangle \lambda}\,.
    \label{eq:visc_2d}
\end{equation}
The viscosity coefficients are:
\begin{equation}
    \lambda^{(e)} = \frac{2\mathfrak p\,\tau_{\mathrm{coll}}}{1+(2\tau_{\mathrm{coll}} B)^2}\,, ~~~~~~ \lambda^{(o)}=\frac{4\mathfrak p\,B\,\tau_{\mathrm{coll}}^2}{1+(2\tau_{\mathrm{coll}} B)^2}\,.
    \label{eq:visc_coeffs_2d}
\end{equation}
Regarding the ''heat'' current $q^\mu$,
\begin{equation}
    q^\mu = -\kappa^{(e)} \partial^\mu T -\kappa^{(o)}u^{\rho}\epsilon_{\rho\lambda\nu}h^{\mu\lambda}\partial^\nu T,
    \label{eq:q_2d}
\end{equation}
with the even and odd thermal conductivity coefficients equal to 
\begin{equation}
    \kappa^{(e)}= \frac{\epsilon+\mathfrak p}{\rho}\frac{n\,\tau_{\mathrm{coll}}}{1+(\tau_{\mathrm{coll}} B)^2}\,, ~~~~~~~~\kappa^{(o)} = \frac{\epsilon+\mathfrak p}{\rho}\frac{\tau_{\mathrm{coll}}^2 B  }{1+(\tau_{\mathrm{coll}} B)^2}\,. 
\end{equation}

The constitutive relations in a frame with observer velocity $v^\mu$ up to the first order in the derivative expansion read
\begin{align}
    \mathcal P_\mu &= \rho \,u_\mu \,,\qquad T_{\mu\nu} = \mathfrak p\, h_{\mu\nu}  -\eta_{\mu\nu\alpha\beta} \sigma^{\alpha\beta}\,,\\
    \tilde{\mathcal E}^\mu &= \left(\epsilon + \frac{\rho}{2}u^2 \right)u^\mu + \mathfrak p \,P^\mu_\nu\,u^\nu -\kappa^{(e)} \partial^\mu T -\kappa^{(o)}u^{\rho}\epsilon_{\rho\lambda\nu}h^{\mu\lambda}\partial^\nu T -\eta^{\mu\nu}_{\alpha\beta} u_\nu\sigma^{\alpha\beta}\,.
\end{align}

\subsubsection{Three-dimensional rotating gas}
Let us now consider the more complicated case of a 3D gas. In order to express the constitutive relations covariantly, we first introduce the \textit{vorticity vectors} with lower index ($B_\mu$) and upper index ($B^\mu$), their norm $|B|$, and the corresponding unit vectors $b_\mu$ and $b^\mu$, as follows:
\begin{equation}
    B_\mu = \frac12u^\rho \epsilon_{\rho\sigma\nu\mu}f^{\sigma\nu}\,,~~B^\mu = h^{\mu\nu}B_\nu\,,~~|B|^2 = B^\lambda B_\lambda\,,~~b_\mu = \frac{B_\mu}{|B|}\,,~~b^\mu = \frac{B^\mu}{|B|}\,.
    \label{eq:vorticity_vector}
\end{equation}
Note that the vectors above are purely spatial, i.e., $\tau_\mu b^\mu = u^\mu b_\mu = 0$. They can be used to form two spatial projection operators, $R^\mu_\nu$ longitudinal to $b^\mu$ and $Q^\mu_\nu$ transverse to $b^\mu$:
\begin{equation}
    R^\mu_\nu = b^\mu b_\nu\,,~~~~~~Q^\mu_\nu = \bar P^\mu_\nu - b^\mu b_\nu\,.
\end{equation}
The viscosity tensor $\eta^{\mu\nu}_{\alpha\beta}$ in the absence of bulk viscosity can be written as a sum of five terms, \cite{Baginskii1965,lifshitz1981physical}
\begin{equation}
    \eta^{\mu\nu}_{\alpha\beta} = \lambda^{(1e)} \eta^{(1e)\mu\nu}_{\alpha\beta}+\lambda^{(2e)} \eta^{(2e)\mu\nu}_{\alpha\beta}+\lambda^{(3e)} \eta^{(3e)\mu\nu}_{\alpha\beta}+\lambda^{(1o)} \eta^{(1o)\mu\nu}_{\alpha\beta}+\lambda^{(2o)} \eta^{(2o)\mu\nu}_{\alpha\beta}\,.
\end{equation}
Explicitly, the tensors have the form:
\begin{equation}
\begin{split}
    \eta^{(1e)\mu\nu}_{\alpha\beta} & = R^{\langle \mu}_{ \alpha}R^{\nu\rangle}_{\beta}\,, ~~~~~~
    \eta^{(2e)\mu\nu}_{\alpha\beta} = R^{\langle \mu}_{\alpha}Q^{\nu\rangle}_{\beta}\,, ~~~~~~
    \eta^{(3e)\mu\nu}_{\alpha\beta} = Q^{\langle \mu}_{ \alpha}Q^{\nu\rangle}_{\beta}\,, \\
    \eta^{(1o)\mu\nu}_{\alpha\beta} & = u^\rho b^\sigma \epsilon_{\rho\sigma\lambda \alpha}R^{\langle \mu}_{\beta}h^{\nu\rangle \lambda}\,, ~~~~~~
    \eta^{(2o)\mu\nu}_{\alpha\beta}  = u^\rho b^\sigma \epsilon_{\rho\sigma\lambda \alpha}Q^{\langle \mu}_{\beta}h^{\nu\rangle \lambda}\,.
\end{split} \label{eq:visc_3d}
\end{equation}
Let us work in cylindrical coordinates $(z,\rho,\phi)$ where the $z$ direction is aligned with the vorticity vector, i.e., $B_\mu dx^\mu = B dz$. In the case of a rotating gas, $B=2\Omega$ and $z$ is the direction of the axis of rotation. Due to magnetic field anisotropy, viscosity coefficients are different for the different directions:
\begin{align}
    \lambda^{(1e)} & = 2\mathfrak p\,\tau_{\mathrm{coll}}\,,~~~~~~ \lambda^{(2e)} = \frac{2\mathfrak p\,\tau_{\mathrm{coll}}}{1+\tau_{\mathrm{coll}}^2 B^2}\,, ~~~~~~\lambda^{(3e)} = \frac{2\mathfrak p\,\tau_{\mathrm{coll}}}{1+4\tau_{\mathrm{coll}}^2 B^2}\,, \\
    \lambda^{(1o)} & = \frac{2\mathfrak p\,B\,\tau_{\mathrm{coll}}^2}{1+\tau_{\mathrm{coll}}^2 B^2}\,,~~~~~~\lambda^{(2o)} = \frac{4\mathfrak p\,B\,\tau_{\mathrm{coll}}^2}{1+4\tau_{\mathrm{coll}}^2 B^2}\,.
\end{align}
The heat current $q^\mu$ reads
\begin{equation}
    q^\mu = -\kappa^{(1e)} R^\mu_\lambda h^{\lambda\nu}\partial_\nu T-\kappa^{(2e)} Q^\mu_\lambda h^{\lambda\nu}\partial_\nu T -\kappa^{(o)}\tau_{\rho}n_\sigma\epsilon^{\rho\sigma\mu\nu}\partial_\nu T\,,
    \label{eq:q_3d}
\end{equation}
with the thermal conductivity coefficients equal to 
\begin{equation}
    \kappa^{(1e)}=\frac{\epsilon+\mathfrak p}{\rho}n\tau_{\mathrm{coll}}\,,~~~~~~~~\kappa^{(2e)}= \frac{\epsilon+\mathfrak p}{\rho}\frac{n\tau_{\mathrm{coll}}}{1+\tau_{\mathrm{coll}}^2 B^2}\,, ~~~~~~~~\kappa^{(o)} = \frac{\epsilon+\mathfrak p}{\rho}\frac{nB\tau_{\mathrm{coll}}^2}{1+\tau_{n\mathrm{coll}}^2 B^2}\,. 
\end{equation}
The values of the viscosity and thermal conductivity coefficients given above agree with the ones found in literature \cite{nakagawa_kinetic_1956,Hartyunyan2016,Mohanty2019,Das2019,Aung2024}. Note that in non-relativistic kinetic theory, bulk viscosity is identically zero when the kinetic energy is fixed in terms of hydrodynamic variables $n$, $T$ and $u^\mu$ as in Eq.~\eqref{eq:eckart} \cite{chapman1990mathematical,nakagawa_kinetic_1956}.

\section{Frame-dependence paradox and its resolution}
\label{sec:invariance}

In the scientific literature, a great deal of controversy about the principle of frame invariance can be found. Let us present the matter of the debate in the language of the NC kinetic theory. Assume for concreteness a two-dimensional gas with zero shear, $\sigma^{\mu\nu}=0$, but non-zero gradient of temperature: $\partial_\mu T\neq 0$. The traditional approach used, e.g., in Ref. \cite{Muller1972}, is to perform the gradient expansion in the frame comoving with the \textit{observer}, rather than in the frame comoving with the \textit{fluid}. In other words, the gradient expansion is performed using $v^\mu$ of the observer, together with the corresponding $h_{\mu\nu}$ and $F_{\mu\nu}$, instead of the fluid-frame quantities $u^\mu$, $g_{\mu\nu}$ and $f_{\mu\nu}$. At the first order, this is equivalent to using the Milne-boost-dependent vorticity scalar $B=\frac12v^\rho \epsilon_{\rho\mu\nu}F^{\mu\nu}$ instead of the Milne-invariant one defined in Eq.~(\ref{eq:vorticity_scalar}). Let us see what effect it has on the constitutive relations. Firstly, for an observer with zero vorticity, $B=0$, the heat current is given by
\begin{equation}
    q_1^\mu = -\tau_{\mathrm{coll}}(2nT) \partial^\mu T\,.
\end{equation}
Now let us change the frame of reference to one rotating with frequency $\Omega$, as in Section \ref{sec:rotate}. This gives a non-zero vorticity $B=2\Omega$, and, following Eq. (\ref{eq:q_2d}), we then have
\begin{equation}
    q_2^\mu = -\frac{\tau_{\mathrm{coll}}(2nT)}{1+4\tau_{\mathrm{coll}}^2 \Omega^2} \partial^\mu T -\frac{2\tau_{\mathrm{coll}}^2\Omega(2nT)}{1+4\tau_{\mathrm{coll}}^2 \Omega^2}u^{\rho}\epsilon_{\rho\lambda\nu}h^{\mu\lambda}\partial^\nu T\,.
\end{equation}
The problem is that $q_1^\mu\neq q_2^\mu$, despite the seemingly covariant notation for $q^\mu$. This apparent contradiction led the author of Ref. \cite{Muller1972} to conclude that constitutive relations depend on the observer's frame of reference. Taking this point of view, however, leads to a paradoxical situation: the heat transport equation in the absence of shear,
\begin{equation}
    \frac{d}{2}n\, u^\mu \partial_\mu T + n T \nabla_\mu u^\mu + \nabla_\mu q^\mu = 0\,,
\end{equation}
appears to change its form under different choices of observer frame, even though \( T \) is a frame-invariant scalar.

The source of the problem, then, is that the usual Newtonian approach to non-relativistic physics implicitly introduces the observer velocity $v^\mu$, which is in principle arbitrary, into the gradient expansion procedure. The way in which $v^\mu$ enters the formulas is difficult to spot in a Newtonian formulation, and only becomes evident in a Newton-Cartan treatment. The paradox can be resolved by replacing $v^\mu$ with an unambiguously defined timelike (in the sense $\tau_\mu v^\mu = 1$) vector field. Since for generic gases it is only possible to define one such field, namely the fluid velocity $u^\mu$, it becomes clear that a truly covariant kinetic theory can be obtained after replacing $v^\mu$ with $u^\mu$ in the constitutive relations, which is equivalent to setting the observer velocity $v^\mu=u^\mu$ in the conventional Newtonian approach. This is also consistent with the solution proposed in Refs. \cite{Woods1983, Band1984,Eu1985}, which suggest that constitutive relations should be calculated in the reference frame corotating with the fluid. 

Our argument rests on the following chain of reasoning:

\begin{itemize}
    \item Coordinate-independence is a foundational principle in both relativistic and non-relativistic physics. In the latter case, the Newton--Cartan framework includes not only diffeomorphism invariance but also Milne boost symmetry as a gauge redundancy.

    \item In the presence of a thermal fluid, the fluid velocity field $u^\mu$ selects a preferred element in the Milne equivalence class, spontaneously breaking this symmetry. This breaking is not an artefact of coordinate choice, but rather introduces fixed background structures—such as $u^\mu$ and $g_{\mu\nu}$—that do not transform under Milne boosts and are physically meaningful. These backgrounds play an essential role in defining thermodynamic and hydrodynamic quantities.

    \item The constitutive relations must be built from Milne-invariant data, even if Milne boost symmetry is spontaneously broken. This is analogous to constructing observables in gauge theories from gauge-invariant combinations of fields. The comoving frame naturally arises as a consequence of this symmetry-breaking pattern, not merely as a pragmatic or ``obvious'' choice.

    \item Thus, our claim is that the Milne gauge structure and its spontaneous breaking by the thermal state are essential ingredients for a conceptually complete and consistent theory of non-relativistic, covariant kinetic theory. This insight appears to have been largely overlooked in previous treatments, which either worked in fixed frames or implicitly broke Milne invariance without recognizing it as a gauge symmetry.
\end{itemize}

We now turn our attention to two distinct approaches aimed at addressing the issue of frame invariance in Newtonian physics through the use of covariant formulations.

Firstly, Matolcsi in \cite{Matolcsi1986} introduced a mathematical framework describing the flat non-relativistic four-dimensional spacetime. This formalism was then used by Matolcsi and Gruber in \cite{Matolcsi1996}, where they developed kinetic theory that they argued to be frame-invariant. Matolcsi’s approach restores frame indifference by eliminating any reference to specific observers or coordinate systems. The underlying affine structure of this spacetime provides a canonical flat, torsion-free covariant derivative defined by directional derivatives, so that parallel transport is path-independent and compatibility conditions uniquely determine the clock form and spatial metric. In this sense, Matolcsi’s construction can be regarded as a particular realization of a Newton–Cartan spacetime with closed clock form and flat connection. While this formulation successfully resolves frame indifference in the flat case, it does not directly address the origin of apparent violations in coordinate-based implementations, nor does it offer a route to incorporating curvature or more general geometric structures. The framework developed here aims to close this gap by extending frame-indifferent formulations to settings where coordinate descriptions are indispensable and where nontrivial Newton–Cartan geometries may arise, thereby enabling applications in both flat and curved non-relativistic spacetimes.

Secondly, Frewer in \cite{Frewer2009} considered the issue of frame invariance in the context of curved manifolds that are NC geometries in our formalism. His solution consists of expressing constitutive equations using tensors called in our notation $u^\mu$ and $g_{\mu\nu}$. While expressing the constitutive equations using these tensors is fundamental to our solution of the paradox as well, Frewer makes no mention of the observer velocity $v^\mu$ or the Milne symmetry. In consequence, the reason behind the failure of the previous formulations remains a mystery. In fact, unlike the present paper, Ref. \cite{Frewer2009} is not concerned with developing a covariant kinetic theory, and consequently a covariant implementation of the gradient expansion procedure is also lacking. 

\section{Discussion}

The findings of this study significantly advance our understanding of the kinetic theory in non-relativistic gravitational settings, facilitated by the NC framework. In both relativistic and non-relativistic settings, physical theories must remain observer-independent, even when spacetime symmetries are broken. In Newton--Cartan geometry, fluids in equilibrium select a preferred frame, spontaneously breaking Milne boost invariance. This breaking introduces fixed background fields, such as the fluid velocity, which do not transform under the broken symmetry. Though the equations remain covariant, these backgrounds encode physically meaningful symmetry breaking, making its effects observable. A consistent use of observer-independent data is pivotal in establishing a robust theoretical foundation for analyzing gases and fluids in gravitational fields. As a consequence we are also able to resolve a long-standing puzzle in kinetic theory that concerns the covariant formulation. Without explicitly enforcing the Milne boost symmetry, and pointing out where the traditional approach to kinetic theory fails to observe it, it is impossible to understand the resolution of the frame-indifference controversy. Previous attempts to solve this problem have missed this symmetry and thus have not been able to provide a clear physical explanation of the paradox. 

Our work not only advances the theoretical understanding of kinetic theory in curved spacetime, but also opens up new avenues for experimental investigation in ultracold atomic systems. The integration of geometric potentials and quantum vortices provides a rich framework for exploring the dynamics of superfluids, and the experimental feasibility of studying these systems in microgravity environments further enhances the potential impact of our findings. Recent experimental advances in Bose-Einstein condensates on the International Space Station allow the creation of ultracold atomic bubbles \cite{aveline_observation_2020,carollo_observation_2022}. This provides a promising possibility to investigate a bubble-trapped superfluid experimentally. Motivated by this experimental progress, there has been a renewed interest in the dynamics of few-body vortices on curved surfaces \cite{bereta_superfluid_2021,caracanhas_superfluid_2022,padavic_vortex-antivortex_2020,xiong_hydrodynamics_2024}, offering different perspectives on the mathematical treatment of point vortex dynamics on curved surfaces.

The covariant formulation of kinetic theory allows one to systematically study transport in non-inertial frames and under the influence of gravity. Helioseismology has unveiled the internal rotation profile of the Sun \cite{thompson_internal_2003}. The inner radiative zone exhibits rigid rotation, whereas the outer convective zone displays a complex differential rotation profile. Separating these two zones is a thin layer known as the solar tachocline, which marks the transition from rigid to differential rotation. Current observations indicate that the tachocline's thickness and position remain relatively stable over time, and it is characterized by strong toroidal magnetic fields. The dynamics of the solar tachocline are a focal point of extensive research \cite{strugarek_dynamics_2023}. Typically, the plasma dynamics within this thin layer are approximated by the shallow water magnetohydrodynamic equations \cite{schecter_shallow-water_2001}. In this approximation, the curvature effects are neglected. Our frame-indifferent kinetic theory approach enables a systematic incorporation of geometric effects into the magnetohydrodynamic  equations and shedding further light on the dynamics of Sun.

\section*{Acknowledgements}\label{sec:ack}

Special thanks to Piotr Witkowski for his contributions and involvement in the early stages of this work. R.B. and P.S. have been supported, in part, by
the Polish National Science Centre (NCN) Sonata Bis
Grant 2019/34/E/ST3/00405. F.P.-B. has received funding
from the Norwegian Financial Mechanism 2014-2021 via the
NCN, POLS Grant 2020/37/K/ST3/03390. P.M. has been supported through the International Max Planck Research School (IMPRS) for Quantum Dynamics and Control hosted at the Max Planck Institute for the Physics of Complex Systems.

\appendix

\section{Gradient expansion}
\label{sec:gradient}

By examining Eq.~(\ref{eq:moments}), the following relation between the $k$-th and the $(k-1)$-th moment of the distribution function can be derived:
\begin{equation}
    \tau_{\alpha_k}\mathcal{I}^{\alpha_1\alpha_2\cdots\alpha_{k-1}\alpha_k}=\mathcal{I}^{\alpha_1\alpha_2\cdots\alpha_{k-1}}\,.
\end{equation}
In other words, the temporal component of the $k$-th moment is equal to the $(k-1)$-th moment. Therefore, we can focus on finding only the spatial components of the moments. We distinguish the purely spatial parts of $\mathcal{I}^{\alpha_1\alpha_2\cdots\alpha_k}$, $\mathcal{K}^{\alpha_1\alpha_2\cdots\alpha_k}$ and $\mathcal{S}^{\alpha_1\alpha_2\cdots\alpha_k}$ by a plain font:
\begin{equation}
\begin{split}
    I^{\alpha_1\alpha_2\cdots\alpha_k} &= \bar{P}^{\alpha_1}_{\alpha_1'}\bar{P}^{\alpha_2}_{\alpha_2'}\cdots\bar{P}^{\alpha_k}_{\alpha_k'} \mathcal{I}^{\alpha'_1\alpha'_2\cdots\alpha'_k}\,, \\
    K^{\alpha_1\alpha_2\cdots\alpha_k} &= \bar{P}^{\alpha_1}_{\alpha_1'}\bar{P}^{\alpha_2}_{\alpha_2'}\cdots\bar{P}^{\alpha_k}_{\alpha_k'} \mathcal{K}^{\alpha'_1\alpha'_2\cdots\alpha'_k}\,,\\
    S^{\alpha_1\alpha_2\cdots\alpha_k} &= \bar{P}^{\alpha_1}_{\alpha_1'}\bar{P}^{\alpha_2}_{\alpha_2'}\cdots\bar{P}^{\alpha_k}_{\alpha_k'} \mathcal{S}^{\alpha'_1\alpha'_2\cdots\alpha'_k}\,.
    \label{eq:I_C_defs}
\end{split}
\end{equation}
Note that the spatial projection used here is the one defined in Eq. (\ref{eq:P_inv}) using $u^\mu$. Applying the projection operators $\bar{P}^{\alpha_1}_{\alpha_1'}\bar{P}^{\alpha_2}_{\alpha_2'}\cdots\bar{P}^{\alpha_k}_{\alpha_k'}$ to both sides of Eq. (\ref{eq:moments_covariant}) gives, in the torsionless case $H_{\mu\nu}=0$,
\begin{equation}
    \bar{P}^{\alpha_1}_{\alpha_1'}\bar{P}^{\alpha_2}_{\alpha_2'}\cdots \bar{P}^{\alpha_k}_{\alpha_k'}\nabla_\mu \mathcal{I}^{\mu\alpha_1'\alpha_2'\cdots\alpha_k'} = K^{\alpha_1\alpha_2\cdots\alpha_k}\,. 
    \label{eq:deriv2_init}
\end{equation}
The projection operators on the left-hand side of Eq.~(\ref{eq:deriv2_init}) can be shifted inside the covariant derivative at the price of generating new terms in the equation, which is described in detail in Appendix~\ref{app:deriv_eom}. In the end, Eq.~\eqref{eq:deriv2_init} takes the form
\begin{align}
    \left(\pounds_u+\nabla_\mu u^\mu\right) I^{\alpha_1\cdots \alpha_k}&+\nabla_\mu I^{\mu\alpha_1\cdots\alpha_k} 
    +\big[h^{\alpha_k\sigma}\left(u^\mu I^{\alpha_1\cdots\alpha_{k-1}}+2I^{\mu\alpha_1\cdots\alpha_{k-1}}\right)+\cdots
    \nn
     &\cdots +h^{\alpha_1\sigma}\left(u^\mu I^{\alpha_2\cdots \alpha_k}+2I^{\mu\alpha_2\cdots \alpha_k}\right)\big]\left(g_{\lambda\sigma}\nabla_\mu u^\lambda\right)
    = K^{\alpha_1\cdots\alpha_k}\,,
    \label{eq:deriv2_fifth}
\end{align}
where the Lie derivative of $I^{\alpha_1\cdots\alpha_k}$ with respect to $u^\mu$ is
\begin{equation}
    \pounds_u I^{\alpha_1\cdots\alpha_k} = u^\mu \partial_\mu I^{\alpha_1\cdots\alpha_k} - I^{\mu\alpha_2\cdots\alpha_k}\partial_\mu u^{\alpha_1} - \cdots - I^{\alpha_1\cdots\alpha_{k-1}\mu}\partial_\mu u^{\alpha_k}\,.
\end{equation}
By splitting $\nabla_\mu u^\lambda$ into its components as in Eqs.~(\ref{eq:nabla_u_decompose})-(\ref{eq:def_theta}), we can write Eq.~\eqref{eq:deriv2_fifth} in a compact form as
\begin{equation}
    T_{(k)}I^{(k)} + D_{(k)}I^{(k+1)}+ E_{(k)}I^{(k-1)} =K^{(k)}\,,~~~~k\geq 0
    \label{eq:moments_general}
\end{equation}
where $I^{(k)}$ and $K^{(k)}$ denote symmetric tensors with $k$ indices defined in Eq. (\ref{eq:I_C_defs}) with the indices left implicit, and we have defined
\begin{align}
    \left[T_{(k)}I^{(k)}\right]^{\alpha_1\alpha_2\cdots\alpha_k} &\equiv \left(\pounds_u+\theta\right)I^{\alpha_1\alpha_2\cdots\alpha_k}+k I^{\mu(\alpha_2 \cdots \alpha_k}\bar{f}^{\,\nu|\alpha_1)}g_{\mu\nu}\,,\\
    \left[D_{(k)}I^{(k+1)}\right]^{\alpha_1\alpha_2\cdots\alpha_k} &\equiv \nabla_\mu I^{\mu \alpha_1\alpha_2\cdots\alpha_k}\,,\\
    \left[E_{(k)}I^{(k-1)}\right]^{\alpha_1\alpha_2\cdots\alpha_k} &\equiv -k I^{(\alpha_2 \cdots \alpha_k}f^{\alpha_1)}\,_{\rho}u^\rho\,,
\end{align}
where
\begin{equation}
    \bar{f}^{\mu\nu}\equiv f^{\mu\nu}+2\sigma^{\mu\nu}+\frac{2}{d}h^{\mu\nu}\theta\,.
    \label{eq:F_general}
\end{equation}
We use the convention that $I^{(-1)}=0$. The notation emphasizes that $T_{(k)}$ is a linear map from the space of symmetric tensors with $(k)$ indices to itself, $T_{(k)}: \{I^{(k)}\}\rightarrow\{I^{(k)}\}$, while $D_{(k)}:\{I^{(k+1)}\}\rightarrow \{I^{(k)}\}$ and $E_{(k)}:\{I^{(k-1)}\}\rightarrow \{I^{(k)}\}$. Physically, $T_{(k)}$ corresponds to a combination of a generalized time derivative and the action of a generalized magnetic field, $D_{(k)}$ is the generalized spatial derivative, while $E_{(k)}$ gives the action of the generalized electric field.

Within the relaxation time approximation, see Eq.~(\ref{eq:tra}), the general equation (\ref{eq:moments_general}) can be further rewritten. Firstly, the equations for $k=0$, $k=1$, and the trace of the equation for $k=2$, can be expressed as
\begin{align}
    &u^\mu\partial_\mu n+n \nabla_\mu  u^\mu=0\,, \label{eq:n}\\
    &h^{\mu\nu}\partial_\mu p -mnf^\nu\,_{\mu}u^\mu+\nabla_\mu \tau^{\mu\nu}=0\,, \label{eq:u}\\
    &\frac{d}{2}nu^\mu\partial_\mu T+ nT\nabla_\mu u^\mu+\tau^{\mu\nu}g_{\rho\mu}\nabla_\nu u^\rho+\nabla_\mu q^\mu=0\,, \label{eq:e}
\end{align}
i.e., they take the form of equations \eqref{eq:eoms_thermo_covariant-a}-\eqref{eq:eoms_thermo_covariant-c}. Secondly, the equations of motion for the out-of-equilibrium moments $\delta I^{\alpha \beta \cdots \gamma}$ take the form:
\begin{align}
    \left[\mathcal{T}_{(2)}\delta I^{(2)} + D_{(2)}\delta I^{(3)}\right]^{\langle\alpha\beta\rangle} & =S^{\langle \alpha\beta\rangle}\,, \label{eq:2nd_mom}\\
   \mathcal{T}_{(k)}\delta I^{(k)} + D_{(k)}\delta I^{(k+1)}+ E_{(k)}\delta I^{(k-1)} & =S^{(k)}\,,~~~~k\geq 3\,, \label{eq:recurs}
\end{align}
where $S^{(k)}$ is the spatial part of the operator defined in Eq.~\eqref{eq:F_covariant} with $k$ indices that we leave implicit, and $\mathcal{T}_{(k)}$ is a combination of $T_{(k)}$ and the collision operator:
\begin{equation}
    \left[\mathcal{T}_{(k)}\delta I^{(k)}\right]^{\alpha_1\alpha_2\cdots\alpha_k}  = \left(\pounds_u +\theta\right)\delta I^{\alpha_1\alpha_2\cdots\alpha_k}+k \delta I^{\mu(\alpha_2 \cdots \alpha_k}\bar{f}^{\nu|\alpha_1)}g_{\mu\nu}+\frac{1}{\tau_{\mathrm{coll}}}\delta I^{\alpha_1\alpha_2\cdots\alpha_k}\,.
    \label{eq:T_def}
\end{equation}

We will now examine the requirements for the hydrodynamic approach to be valid. As discussed in the main text, the condition is that the frequencies of out-of-equilibrium perturbations are much smaller than the relaxation rate $1/\tau_{\mathrm{coll}}$. What complicates the matter is the fact that there are multiple frequency scales that can be defined in this context. To find them, we first introduce a set of operators: $\{\pounds_u, \sigma^{\mu\alpha}g_{\mu\nu},\theta,\frac{T}{m}h^{\mu\nu}\nabla_\mu\nabla_\nu\}$. Then, we can formally define the set of characteristic frequencies of the out-of-equilibrium perturbations $\{\omega,\sigma,\theta,\omega_T^2\}$ as the maximal magnitudes of the eigenvalues of the above-defined operators acting on $\delta \mathcal{I}^{(k)}$. In the hydrodynamic regime, we require $\mathrm{max}\{\omega,\sigma,\theta,\omega_T\}\ll1/\tau_{\mathrm{coll}}$. This brings about two simplifications: one of them is that the terms containing $\pounds_u$, $\theta$ and $\sigma^{\mu\nu}$ in Eq.~(\ref{eq:T_def}) are subleading with respect to $\frac{1}{\tau_{\mathrm{coll}}}\delta I^{(k)}$; the other one is that on dimensional grounds we can estimate $\delta I^{(k+1)}\sim \mathrm{max}\{u_{\mathrm{char}},\sqrt{T/m}\}\delta I^{(k)}$, where $u_{\mathrm{char}}$ stands for the magnitude of velocity perturbation out of equilibrium, so that $D_{(k)}\delta I^{(k+1)}\sim (\sigma \delta I^{(k)},\theta \delta I^{(k)},\omega_T\delta I^{(k)})\ll 1/\tau_{\mathrm{coll}}\delta I^{(k)}$, and consequently in Eq.~(\ref{eq:recurs}) the gradient term $D_{(k)}\delta I^{(k+1)}$ is subleading. This is the essence of the gradient expansion: all the out-of-equilibrium terms are assumed to be subleading, such that an expansion in powers of $\tau_{\mathrm{coll}}\omega,\tau_{\mathrm{coll}}\sigma,\tau_{\mathrm{coll}}\theta,\tau_{\mathrm{coll}}\omega_T$ can be carried out.

The assumptions discussed above significantly simplify the calculations of the first-order gradient corrections. Firstly,
\begin{equation}
    \left[\mathcal{T}_{(k)}\delta I^{(k)}\right]^{\alpha_1\alpha_2\cdots\alpha_k} \approx \frac{1}{\tau_{\mathrm{coll}}}\delta I^{\alpha_1\alpha_2\cdots\alpha_k}-k \delta I^{\mu(\alpha_2 \cdots \alpha_k}f^{\alpha_1)\nu}g_{\mu\nu}\,,
    \label{eq:T_def_2}
\end{equation}
so that $\mathcal{T}_{(k)}$ can be represented as a tensor with $k$ lower and $k$ upper indices rather than as a differential operator. Secondly, in Eq. (\ref{eq:2nd_mom}) the term $D_{(2)}\delta I^{(3)}$ is at least second order in gradients and can thus be neglected, and consequently
\begin{equation}
    \tau^{\mu\nu}= m\delta I^{\mu\nu} = m\left[\mathcal{T}_{(2)}^{-1}S^{(2)}\right]^{\langle \mu \nu \rangle }\,.
    \label{eq:tau_formula}
\end{equation}
Using the definitions in Eqs. (\ref{eq:I_C_defs}) and (\ref{eq:F_covariant}) one can calculate (the details of the calculation are relegated to Appendix \ref{app:deriv_1storder}): 
\begin{equation}
    S^{\alpha\beta}= -\frac{2 n T}{m} h^{\rho\langle \alpha }\nabla_\rho u^{\beta\rangle} = -\frac{2nT}{m}\sigma^{\alpha\beta}\,.
\end{equation}
Analogously, from Eq. (\ref{eq:recurs}) we can deduce that up to the first order in gradients 
\begin{equation}
    \delta I^{(3)} = \mathcal{T}_{(3)}^{-1}\left(S^{(3)}-E_{(3)}\delta I^{(2)}\right)\,.
    \label{eq:I3_formula}
\end{equation}
In Appendix \ref{app:deriv_1storder} we calculate
\begin{equation}
    S^{\alpha\beta\gamma} = -\frac{3nT}{m^2}h^{(\alpha\beta}\partial^{\gamma)} T\,.
    \label{eq:F3_formula}
\end{equation}

Following Eq.~(\ref{eq:tau_formula}), the stress tensor $\tau^{\mu\nu}$ can be obtained by inverting the operators $\mathcal{T}_{(2)}$. In the hydrodynamic regime, $\mathcal{T}_{(2)}$ can be treated as a four-rank tensor such that
\begin{equation}
    mS^{\alpha\beta} = \mathcal{T}^{\alpha\beta}_{~~~\mu\nu}\tau^{\mu\nu}\,.
\end{equation}
Since vorticity $2\nabla^{[\mu}u^{\nu]}=f^{\mu\nu}$ can be non-zero in equilibrium, we take the vorticity scalar defined in Eq.~(\ref{eq:vorticity_scalar}) to be a zeroth-order quantity that has to be taken into account. In two dimensions, we will choose to do calculations in polar coordinates $(\rho,\phi)$. In the basis $\left\{\tau^{\rho\rho},\tau^{\rho\phi},\tau^{\phi\phi}\right\}$,
\begin{equation}
     \mathcal{T}^{\alpha\beta}_{~~~\mu\nu} = \begin{pmatrix}
        \frac{1}{\tau_{\mathrm{coll}}} & -2\rho B & 0 \\
        B/\rho & \frac{1}{\tau_{\mathrm{coll}}} & -\rho B \\
        0&2B/\rho & \frac{1}{\tau_{\mathrm{coll}}}
    \end{pmatrix}\,.
\end{equation}
Changing the basis to $\left\{\mathrm{tr}[\tau]/2,\tau^{\langle \rho\rho \rangle},\tau^{\rho\phi}\right\}$,
\begin{equation}
    \mathcal{T}^{\alpha\beta}_{~~~\mu\nu} = \begin{pmatrix}
        \frac{1}{\tau_{\mathrm{coll}}} & 0 & 0 \\
         0 & \frac{1}{\tau_{\mathrm{coll}}} & -2\rho B \\
        0 &2B/\rho & \frac{1}{\tau_{\mathrm{coll}}}
    \end{pmatrix}\,.
\end{equation}
Inverting this matrix, we obtain
\begin{equation}
\begin{split}
    \frac{1}{2nT}\tau^{\langle \rho\rho\rangle} & = -\frac{\tau_{\mathrm{coll}}}{1+4\tau_{\mathrm{coll}}^2 B^2}\sigma^{\rho\rho}-\frac{2\tau_{\mathrm{coll}}^2B}{1+4\tau_{\mathrm{coll}}^2 B^2}\rho\sigma^{\rho\phi}\,,\\ 
    \frac{1}{2nT}\rho\tau^{\langle \rho\phi\rangle} & = -\frac{\tau_{\mathrm{coll}}}{1+4\tau_{\mathrm{coll}}^2 B^2}\rho\sigma^{ \rho\phi}+\frac{2\tau_{\mathrm{coll}}^2B}{1+4\tau_{\mathrm{coll}}^2 B^2}\sigma^{\rho\rho}\,.
\end{split} \label{eq:tau_2d_cartesian}
\end{equation}
After comparing Eq.~(\ref{eq:tau_2d_cartesian}) with the definitions in Eq.~(\ref{eq:tau_2d}) and~(\ref{eq:visc_2d}), we can read off the viscosity coefficients, which are presented in Eq.~(\ref{eq:visc_coeffs_2d}).

In three dimensions, we work in the cylindrical coordinate system $(z,\rho,\phi)$, where the $z$ direction is aligned with the vorticity vector defined in Eq.~(\ref{eq:vorticity_vector}), i.e., $B_\mu dx^\mu = B dz$. In the basis $\left\{\mathrm{tr}[\tau]/2,(\tau^{ \rho\rho }-\rho^2\tau^{\phi\phi})/2,\tau^{\rho\phi},\tau^{\rho z},\tau^{\phi z}, \tau^{\langle zz\rangle}\right\}$,
\begin{equation}
    \mathcal{T}^{\alpha\beta}_{~~~\mu\nu} = \begin{pmatrix}
        \frac{1}{\tau_{\mathrm{coll}}} & 0 & 0&0 & 0& 0\\
        0 & \frac{1}{\tau_{\mathrm{coll}}} & -2\rho B& 0& 0&0 \\
        0& 2B/\rho& \frac{1}{\tau_{\mathrm{coll}}}&0 &0 &0 \\
        0&0 &0 &\frac{1}{\tau_{\mathrm{coll}}} & -\rho B& 0\\
        0&0 & 0&B/\rho&\frac{1}{\tau_{\mathrm{coll}}} &0 \\
       0 &0 & 0& 0&0 &\frac{1}{\tau_{\mathrm{coll}}}
    \end{pmatrix}\,.
\end{equation}
Inverting this matrix we obtain 
\begin{equation} 
\begin{split}
    \frac{1}{2nT}\frac{\tau^{\langle \rho\rho\rangle}-\rho^2\tau^{\langle \phi\phi\rangle}}{2} & = -\frac{\tau_{\mathrm{coll}}}{1+4\tau_{\mathrm{coll}}^2 B^2}\frac{\sigma^{\rho\rho}-\rho^2\sigma^{ \phi\phi}}{2}-\frac{2\tau_{\mathrm{coll}}^2B}{1+4\tau_{\mathrm{coll}}^2 B^2}\rho\sigma^{ \rho\phi}\,,\\ 
    \frac{1}{2nT}\rho \tau^{\langle\rho\phi\rangle} & = -\frac{\tau_{\mathrm{coll}}}{1+4\tau_{\mathrm{coll}}^2 B^2}\rho\sigma^{ \rho\phi}+\frac{2\tau_{\mathrm{coll}}^2B}{1+4\tau_{\mathrm{coll}}^2 B^2}\frac{\sigma^{ \rho\rho}-\rho^2\sigma^{ \phi\phi}}{2}\,,\\ 
    \frac{1}{2nT}\tau^{\langle\rho z\rangle} & = -\frac{\tau_{\mathrm{coll}}}{1+\tau_{\mathrm{coll}}^2 B^2}\sigma^{ \rho z}-\frac{\tau_{\mathrm{coll}}^2B}{1+\tau_{\mathrm{coll}}^2 B^2}\rho\sigma^{\phi z}\,,\\ 
    \frac{1}{2nT}\rho\tau^{\langle \phi z\rangle} & = -\frac{\tau_{\mathrm{coll}}}{1+\tau_{\mathrm{coll}}^2 B^2}\rho\sigma^{ \phi z}+\frac{\tau_{\mathrm{coll}}^2B}{1+\tau_{\mathrm{coll}}^2 B^2}\sigma^{\rho z}\,,\\
    \frac{1}{2nT}\tau^{\langle zz\rangle} & = -\tau_{\mathrm{coll}}\sigma^{ zz }\,. 
\end{split}    \label{eq:tau_3d_cartesian}
\end{equation}

Analogously, to calculate $q^\mu$, one needs to invert the operator $\mathcal{T}_{(3)}$, as seen from the definition $q^\mu = \frac{m}{2} g_{\alpha\beta}\delta I^{\alpha\beta\mu}$ and Eq. (\ref{eq:I3_formula}). In two dimensions, in the basis $\left\{\delta I^{\rho\rho\rho},\delta I^{\rho\rho\phi},\delta I^{\rho\phi\phi},\delta I^{\phi\phi\phi}\right\}$, $\mathcal{T}_{(3)}$ reads
\begin{equation}
    \mathcal{T}^{\alpha\beta\gamma}_{~~~~~\mu\nu\rho} = \begin{pmatrix}
        \frac{1}{\tau_{\mathrm{coll}}} & -3\rho B &  0&0\\
        B/\rho & \frac{1}{\tau_{\mathrm{coll}}} & -2\rho B & 0\\
         0&2B/\rho & \frac{1}{\tau_{\mathrm{coll}}} & -\rho B \\
         0& 0&3B/\rho& \frac{1}{\tau_{\mathrm{coll}}}
    \end{pmatrix}\,.
\end{equation}
Inverting this matrix and contracting with $g_{\beta\gamma}$, one recovers Eq. (\ref{eq:q_2d}). Similarly, in three dimensions, in the basis $\left\{\delta I^{\rho\rho\rho},\delta I^{\rho\rho\phi},\delta I^{\rho\phi\phi},\delta I^{\phi\phi\phi},\delta I^{\rho\rho z},\delta I^{\rho \phi z},\delta I^{\phi \phi z},\delta I^{\rho zz},\delta I^{\phi zz},\delta I^{zzz}\right\}$,
\begin{equation}
    \mathcal{T}^{\alpha\beta\gamma}_{~~~~~\mu\nu\rho} = \begin{pmatrix}
        \frac{1}{\tau_{\mathrm{coll}}} & -3\rho B & 0 & 0& &0 & 0& 0&0 &0 \\
        B/\rho & \frac{1}{\tau_{\mathrm{coll}}} & -2\rho B &0 &0 &0 &0 & 0&0 &0 \\
         0&2B/\rho & \frac{1}{\tau_{\mathrm{coll}}} & -\rho B & 0& 0& 0& 0&0 &0 \\
         0&0 &3B/\rho& \frac{1}{\tau_{\mathrm{coll}}} & 0&0 &0 &0 & 0& 0\\
          0&0 &0 &0 & \frac{1}{\tau_{\mathrm{coll}}} & -2\rho B &0 &0 & 0&0  \\
          0&0 &0 &0 & B/\rho & \frac{1}{\tau_{\mathrm{coll}}} &  -\rho B &0 & 0&0  \\
          0& 0& 0&0 & 0 & 2B/\rho & \frac{1}{\tau_{\mathrm{coll}}} &0 &0 &0  \\
           0&0 &0 &0 &0 &0 &0 &\frac{1}{\tau_{\mathrm{coll}}} & -\rho B &0 \\
           0&0 &0 &0 &0 &0 &0 & B/\rho & \frac{1}{\tau_{\mathrm{coll}}} &0 \\
           0&0 & 0& 0&0 &0 &0 &0 &0 & \frac{1}{\tau_{\mathrm{coll}}}
    \end{pmatrix}.
\end{equation}
Inverting this matrix and contracting with $g_{\beta\gamma}$, one recovers Eq. (\ref{eq:q_3d}).

\section{Detailed derivations}

\subsection{Noether charge}
\label{app:noether}

Under an arbitrary infinitesimal path deformation \(x^\mu \rightarrow x^\mu+\tilde{\xi}^\mu\), the on-shell variation of the Lagrangian is 
\begin{align}
    \delta_{\mathrm{on}} \mathcal{L} = \tilde{\xi}^\mu \frac{\partial\mathcal{L}}{\partial x^\mu} + \dot{\tilde{\xi}}^\mu\frac{\partial \mathcal{L}}{\partial \dot x^\mu}=\frac{d}{d\lambda}\left(\tilde{\xi}^\mu\frac{\partial \mathcal{L}}{\partial \dot x^\mu}\right)\,,
    \label{eq:var-of-L-in-appx-1}
\end{align}
where we have used the Euler--Lagrange equations to get the last equality. On the other hand, the Lagrangian for a point particle in NC geometry is given by Eq.~\eqref{eq:non-rel-lagran1} as
\begin{align}
    \mathcal{L}=\frac{m}{2N}h_{\alpha\beta}\dot x^\alpha \dot x^\beta + mm_\alpha \dot x^\alpha\,,
    \label{app:non-rel-lagran1}
\end{align}
where \(N=\tau_\rho \dot x^\rho\). The variation of \(\mathcal{L}\) under the shift \(x^\mu \rightarrow x^\mu+\tilde{\xi}^\mu\) can be calculated explicitly as 
\begin{align}
    \delta \mathcal{L} =&\frac{m}{N}h_{\alpha\beta }\dot x^\alpha \dot x^\rho \partial_\rho \tilde{\xi}^\beta + \frac{m}{2N}\dot x^\alpha\dot x^\beta \tilde{\xi}^\mu \partial_\mu h_{\alpha\beta} 
    -\frac{m}{2N^2}h_{\alpha\beta}\dot x^\alpha \dot x^\beta\lb\tau_\mu \dot x^\rho \partial_\rho  \tilde{\xi}^\mu + \dot x^\mu \tilde{\xi}^\rho\partial_\rho \tau_\mu\rb\nn
    &+mm_{\alpha}\dot x^\rho \partial_\rho \tilde{\xi}^\alpha + m \dot x^\alpha \tilde{\xi}^\mu \partial_\mu m_\alpha 
    \nn
    =&\frac{m}{2N} \dot x^\alpha \dot x^\beta \lb \tilde{\xi}^\mu \partial_\mu h_{\alpha\beta} + 2 h_{\alpha\mu} \partial_\beta \tilde{\xi}^\mu\rb 
    -\frac{m}{2N^2}h_{\alpha\beta}\dot x^\alpha \dot x^\beta \dot x^\mu  \lb\tilde{\xi}^\rho\partial_\rho\tau_\mu + \tau_\rho\partial_\mu \tilde{\xi}^\rho \rb 
    \nn
    &+ m\dot x^\alpha \lb \tilde{\xi}^\mu \partial_\mu m_\alpha + m_\mu \partial_\alpha \tilde{\xi}^\mu\rb 
    \nn
    =& \frac{m}{2N} \dot x^\alpha \dot x^\beta \pounds_{\tilde{\xi}} h_{\alpha\beta} -\frac{m}{2N^2}h_{\alpha\beta}\dot x^\alpha \dot x^\beta \dot x^\mu \pounds_{\tilde{\xi}} \tau_\mu + m \dot x^\alpha \pounds_{\tilde{\xi}} m_\alpha\,,
    \label{eq:var-of-L-in-appx-2}
\end{align}
Let us denote all the transformations as \(\chi=\lb  \tilde{\xi}^\mu,\psi_\mu,\Lambda\rb\), where \(\tilde{\xi}^\mu\) is an infinitesimal coordinate reparameterization, \(\psi_\mu\) is a Milne boost, and \(\Lambda\) is a \(U(1)\) gauge transformation. If the parameters satisfy the Killing conditions in Eq.~(\ref{eq:killing_conds}), the transformation \(\chi\) leaves the action invariant up to a boundary term, and the variation of the Lagrangian under \(\chi\) is called the symmetry variation~\cite{vujanovic_group-variational_1970,djukic_procedure_1973,Sarlet_Generalizations_1981,manton_topological_2004}. If we denote by \(\chi_K\) the set of Killing parameters, then we have 
\begin{align}
    & \delta_{\chi_K} \tau_\mu = \pounds_{\xi_K} \tau_\mu =0\,,\label{eq:app-var-tau}\\
    &\delta_{\chi_K} h_{\alpha\beta}=\pounds_{\xi_K} h_{\alpha\beta} -\psi^K_\alpha\tau_\beta -\psi^K_\beta\tau_\alpha=0
    \nn
    &\Rightarrow \pounds_{\xi_K} h_{\alpha\beta} =\psi^K_\alpha\tau_\beta +\psi^K_\beta\tau_\alpha\,,
    \label{eq:app-var-h}\\
    &\delta_{\chi_K} m_{\alpha}=\pounds_{\xi_K} m_{\alpha}+\psi^K_\alpha +\partial_\alpha \Lambda_K=0\nn
    &\Rightarrow \pounds_{\xi_K} m_{\alpha}= -\psi^K_\alpha - \partial_\alpha \Lambda_K\,,
     \label{eq:app-var-m}
\end{align}
For a symmetry variation, we can use Eqs.~\eqref{eq:app-var-tau},~\eqref{eq:app-var-h} and~\eqref{eq:app-var-m} to simplify Eq.~\eqref{eq:var-of-L-in-appx-2}. Thus, the symmetry variation of \(\mathcal{L}\) reads
\begin{align}
    \delta_{\mathrm{sy}} \mathcal{L}=-m\frac{d}{d\lambda}\Lambda_K\,,
    \label{eq:var-of-L-in-appx-3}
\end{align}
If the arbitrary coordinate reparametrization \(\tilde{\xi}^\mu\) in the on-shell variation in Eq.~\eqref{eq:var-of-L-in-appx-1} is replaced by the Killing transformation \(\xi^\mu = \xi^\mu_K\), then the left-hand sides of both the transformations are equal~\cite{banados_short_2016}. Equating the on-shell and the symmetry variation in  Eqs.~\eqref{eq:var-of-L-in-appx-1} and~\eqref{eq:var-of-L-in-appx-3} we get
\begin{align}
    &\frac{d}{d\lambda}\lb\xi_K^\mu\frac{\partial \mathcal{L}}{\partial \dot x^\mu} + m\Lambda_K\rb=0~\nn
    &\Rightarrow \frac{dQ_K}{d\lambda}=0\,,
\end{align}
where \(Q_K=\xi_K^\mu\frac{\partial \mathcal{L}}{\partial \dot x^\mu} + m\Lambda_K=\xi_K^\mu\pi_\mu +m\Lambda_K\) and \(\pi_\mu=\frac{\partial \mathcal{L}}{\partial \dot x^\mu}\) is the conjugate momentum of the particle. This calculation shows that \(Q_K\) is a conserved quantity known as the Noether charge.
\subsection{Derivation of Eq.~(\ref{eq:LX_f})}
\label{app:deriv_LX_Q}
The collisional invariant in Eq.~(\ref{eq:Q_general}) is
\begin{align}
    Q_0=\xi^\mu\pi_\mu+m\Lambda\,,
\end{align}
where $\xi^\mu$ is a vector, $\pi_\mu$ is the canonical momentum, and $\Lambda$ is a gauge parameter. We define
\begin{align}
    \bar h_{\alpha\beta}&=h_{\alpha\beta}+\tau_\alpha m_\beta +\tau_\beta m_\alpha\,,\\
    \hat{v}^\mu &= v^\mu-h^{\mu\sigma}m_\sigma\,.
\end{align} 
It is straightforward to verify that $\bar h_{\alpha\beta}$ and $\hat{v}^\mu$ are Milne-invariant. An explicit calculation shows that
\begin{equation}
\begin{split} \label{eq:app_noether_aux}
    \Gamma^{\mu}_{\alpha\beta} =&\, \hat{v}^\mu \partial_{\beta}\tau_\alpha + \frac{1}{2}h^{\mu\sigma}\lb \partial_\alpha \bar h_{\beta\sigma} + \partial_{\beta}\bar h_{\alpha\sigma} - \partial_\sigma \bar h_{\alpha\beta}\rb +\frac{1}{2}h^{\mu\sigma}\lb m_\sigma H_{\beta\alpha}+m_{\alpha} H_{\sigma\beta}+m_{\beta}H_{\sigma\alpha}\rb\,,\\
    \hat{v}^\mu \bar{h}_{\mu\nu}=&\left(2v^\sigma m_\sigma - h^{\gamma\sigma}m_\gamma m_\sigma\right)\tau_\nu \,,\\
    h^{\mu\sigma}\bar{h}_{\sigma\nu}=&\,\delta^{\mu}_\nu +\left(h^{\mu\sigma}m_\sigma-v^\mu\right) \tau_\nu\,.
\end{split}
\end{equation}
Now, the canonical momentum $\pi_\mu$ can be rewritten as
\begin{align}
    \pi_\mu=-\frac{1}{2m}\Bar{h}_{\alpha\beta} p^\alpha p^\beta \tau_\mu + \bar h_{\mu\alpha}p^\alpha\,,
\end{align}
where $p^\alpha$ is the kinematic momentum satisfying the constraint $\tau_\mu p^\mu=m$. Acting with the Liouville operator on the collisional invariant gives
\begin{align}
   m X_L[Q_0]=&mp^\gamma\partial_\gamma Q_0-\lb\Gamma^\gamma_{\alpha\beta}p^\alpha p^\beta+\frac{p^2}{2m}H^{\gamma}{}_\rho p^\rho\rb\frac{\partial Q_0}{\partial p^\gamma}\nn
    =&p^\gamma\lb\xi^\mu\partial_\gamma \pi_\mu +\pi_\mu \partial_\gamma \xi^\mu + m\partial_\gamma \Lambda\rb-\lb\Gamma^\gamma_{\alpha\beta}p^\alpha p^\beta+\frac{p^2}{2m}H^{\gamma}{}_\rho p^\rho\rb \xi^\mu \frac{\partial \pi_\mu}{\partial p^\gamma}\nn
    =&p^\gamma \xi^\mu \left[-\frac{1}{2m}\lb\partial_\gamma \bar h_{\alpha\beta}\rb p^\alpha p^\beta \tau_\mu -\frac{1}{2m}\bar h_{\alpha\beta}p^\alpha p^\beta \partial_\gamma \tau_\mu +\lb \partial_\gamma \bar h_{\mu\alpha}\rb p^\alpha\right]\nn
    &+p^\gamma \lb -\frac{1}{2m}\Bar{h}_{\alpha\beta} p^\alpha p^\beta \tau_\mu + \bar h_{\mu\alpha}p^\alpha \rb \partial_\gamma \xi^\mu + m p^\gamma \partial_\gamma \Lambda \nn
    &-\lb\Gamma^\gamma_{\alpha\beta}p^\alpha p^\beta+\frac{p^2}{2m}H^{\gamma}{}_\rho p^\rho\rb \xi^\mu \lb -\frac{1}{m}\bar h_{\gamma\sigma}p^\sigma\tau_\mu + \bar h_{\mu\gamma}\rb\,.
\end{align}
We can further rewrite it as
\begin{align}
   m X_L[Q_0]=& p^\alpha p^\beta \xi^\mu\lb -\frac{1}{2m}\tau_\mu p^\gamma \partial_\gamma\bar h_{\alpha\beta}  + \partial_\beta \bar h_{\mu\alpha} -\lb\Gamma^\gamma_{\alpha\beta}+\frac{1}{2m}h_{\alpha\beta}H^{\gamma}{}_\rho p^\rho\rb \lb -\frac{1}{m}\bar h_{\gamma\sigma}p^\sigma\tau_\mu + \bar h_{\mu\gamma}\rb \rb
    \nn
    &-\frac{1}{2m}\bar h_{\alpha\beta}  p^\alpha p^\beta p^\gamma\mathcal{L}_{\xi}\tau_\gamma + \frac{1}{2m}\bar h_{\alpha\beta} p^\alpha p^\beta p^\gamma \xi^\mu H_{\mu\gamma} +\bar h_{\mu\alpha} p^\alpha p^\beta \partial_\beta \xi^\mu +m p^\gamma \partial_\gamma  \Lambda\,.
    \label{eq:a14}
\end{align}
Using the identities in Eq.~(\ref{eq:app_noether_aux}), as well as $\tau_\rho \lb \frac{p^\rho\tau_\mu}{m}-\delta^{\rho}_{\mu}\rb=0$, we have explicitly
\begin{align}
    \begin{split}
    &\Gamma^\gamma_{\alpha\beta} \bar h_{\gamma\rho}\frac{p^\rho\tau_\mu}{m} -  \Gamma^\gamma_{\alpha\beta}\bar h_{\mu\gamma}
    = \Gamma^\gamma_{\alpha\beta} \bar h_{\gamma\rho}\lb \frac{p^\rho\tau_\mu}{m}-\delta^{\rho}_{\mu}\rb\\
    &= \frac12\left[\lb\partial_\alpha \bar h_{\beta\rho}+\partial_\beta \bar h_{\alpha\rho}-\partial_\rho \bar h_{\alpha\beta}\rb+\lb m_\alpha H_{\rho\beta}+m_\beta H_{\rho\alpha}+m_\rho H_{\beta\alpha}\rb\right]\lb \frac{p^\rho\tau_\mu}{m}-\delta^{\rho}_{\mu}\rb \,, \label{eq:a15}
    \end{split}\\
    &\frac{1}{2m}h_{\alpha\beta}H^\gamma{}_\rho p^\rho \bar h_{\gamma\sigma}\lb \frac{p^\sigma\tau_\mu}{m}-\delta^{\sigma}_{\mu}\rb=\frac{1}{2m}h_{\alpha\beta}H_{\sigma\rho} p^\rho \lb \frac{p^\sigma\tau_\mu}{m}-\delta^{\sigma}_{\mu}\rb=-\frac{1}{2m}h_{\alpha\beta}H_{\mu\rho}p^\rho\,, \\
    & \frac{1}{2m}\bar h_{\alpha\beta} p^\alpha p^\beta p^\gamma \xi^\mu H_{\mu\gamma} = \frac{1}{2m} h_{\alpha\beta} p^\alpha p^\beta p^\gamma \xi^\mu H_{\mu\gamma}+ m_\alpha p^\alpha p^\gamma \xi^\mu H_{\mu\gamma}\,.
\end{align}
Let us first focus on the terms containing $H_{\mu\nu}$. After expanding $\lb \frac{p^\rho\tau_\mu}{m}-\delta^{\rho}_{\mu}\rb$ in Eq.~(\ref{eq:a15}), there are nine such terms. Four of them vanish by the virtue of the antisymmetricity of $H_{\mu\nu}$, which implies $p^\alpha p^\beta H_{\beta\alpha}=p^\alpha p^\rho H_{\rho\alpha}=p^\rho p^\beta H_{\rho\beta}=0$, and the remaining five mutually cancel, so that in the end the tensor $H_{\mu\nu}$ disappears from Eq.~(\ref{eq:a14}). Furthermore, after expanding the terms containing derivatives of $\bar h_{\mu\nu}$, most of them cancel out as well. In the end, Eq.~\eqref{eq:a14} simplifies to
\begin{align}
    mX_L[Q_0]=&\frac{1}{2}p^\alpha p^\beta \xi^\mu \partial_\mu \bar h_{\alpha\beta} + \bar h_{\mu\alpha} p^\alpha p^\beta \partial_\beta \xi^\mu + p^\alpha \tau_\alpha p^\beta \partial_\beta  \Lambda -\frac{1}{2m}\bar h_{\alpha\beta}  p^\alpha p^\beta p^\gamma\mathcal{L}_{\xi}\tau_\gamma\nn
    =&\frac{1}{2}p^\alpha p^\beta \lb \xi^\mu \partial_\mu \bar h_{\alpha\beta} +\bar h_{\alpha\mu}\partial_\beta \xi^\mu +\bar h_{\beta\mu}\partial_\alpha \xi^\mu  +\tau_\alpha\partial_\beta \Lambda+\tau_\beta\partial_\alpha\Lambda\rb-\frac{1}{2m}\bar h_{\alpha\beta}  p^\alpha p^\beta p^\gamma\mathcal{L}_{\xi}\tau_\gamma \nn
    =&\frac{1}{2}p^\alpha p^\beta \delta_\chi \bar h_{\alpha\beta} -\frac{1}{2m}\bar h_{\alpha\beta}  p^\alpha p^\beta p^\gamma \delta_{\chi}\tau_\gamma\,.
    \label{eq:a16}
\end{align}
To find the formula for $\delta_\chi \bar h_{\alpha\beta}$, we have used Eqs.~(\ref{eq:var_of_tau})-(\ref{eq:var_of_A}). Now, since $\bar h_{\alpha\beta}$ is Milne-invariant, setting $v^\mu=u^\mu$ we obtain the equality
\begin{equation}
    \bar h_{\alpha\beta} = g_{\alpha\beta}+2\tau_{(\alpha}A_{\beta)}\,.
    \label{eq:a17}
\end{equation}
From the definition of $u^\mu$, Eq.~(\ref{eq:T-and-u}), we have
\begin{equation}
    \delta_\chi u^\mu = - u^\mu u^\sigma \delta_\chi \tau_\sigma\,,
\end{equation}
which implies
\begin{align}
    u^\alpha \delta_{\chi} g_{\alpha\beta} & = \delta_{\chi}\lb u^\alpha g_{\alpha \beta}\rb-g_{\alpha\beta}\delta_\chi u^\alpha =g_{\alpha\beta}u^\alpha u^\sigma \delta_\chi \tau_\sigma=0\,,\\
    \delta_\chi \bar P^\mu_\alpha & = -u^\mu \bar P^\sigma_\alpha \delta_\chi \tau_\sigma\,,
\end{align}
which in turn can be used to prove that
\begin{equation}
\begin{split}
    \delta_\chi g_{\alpha\beta} & = \lb \delta^\mu_\alpha-u^\mu \tau_\alpha\rb \delta_\chi g_{\mu\beta}= h^{\mu\nu} g_{\alpha\nu}\delta_\chi g_{\mu\beta}=g_{\alpha\nu}\lb -u^\nu \bar P^\sigma_\beta \delta_{\chi}\tau_\sigma \rb-g_{\alpha\nu}g_{\beta\mu}\delta_\chi h^{\mu\nu}\\
    & =-g_{\alpha\nu}g_{\beta\mu}\delta_\chi h^{\mu\nu}\,.
    \label{eq:a18}
\end{split}
\end{equation}
Taking together Eqs.~(\ref{eq:a16}), (\ref{eq:a17}), and (\ref{eq:a18}), gives
\begin{equation}
    mX_L[Q_0]=p^\beta \delta_{\chi} A_\beta-\frac{1}{2}p^\alpha p^\beta g_{\alpha\nu}g_{\beta\mu}\delta_\chi h^{\mu\nu} -\frac{1}{2m} g_{\alpha\beta}  p^\alpha p^\beta p^\gamma \delta_{\chi}\tau_\gamma\,.
\end{equation}

\subsection{Covariant derivative of \texorpdfstring{$\mathcal{I}^{\alpha_1 \alpha_2\cdots\alpha_n}$}{TEXT}}
\label{app:deriv_volume}

For the purposes of this Appendix, we define 
\begin{equation}
    \sigma_{d}=\sigma_{d+1}\delta(\tau_\rho p^\rho-m)\,.
\end{equation}
The covariant derivative acts on the $n$-th moment of the distribution function as 
\begin{equation}
\begin{split}
    \bar\nabla_{\mu}\mathcal{I}^{\mu \alpha_2\cdots\alpha_n}=& \frac{1}{m^n}\bar\nabla_\mu \int   p^\mu p^{\alpha_2} \cdots p^{\alpha_n} f\sigma_d\\
    =&\bar\Gamma^\mu_{\lambda\mu}\mathcal{I}^{\lambda \alpha_2\cdots\alpha_n}+\cdots+\bar\Gamma^{\alpha_n}_{\lambda\mu}\mathcal{I}^{\mu \alpha_2\cdots\lambda}+\frac{1}{m^n}\int   p^\mu p^{\alpha_2} \cdots p^{\alpha_n} \bar\nabla_\mu \left(f\sigma_d\right)\,.
    \label{eq:div_j_1st}
\end{split}
\end{equation}
Now,
\begin{align}
    \bar\nabla_\mu(f\sigma_d) &= \partial_\mu f\sigma_d + f\bar\nabla_\mu\sigma_d \\
    &= \partial_\mu f\sigma_d + f\bar\Gamma^\rho_{\rho\mu}\sigma_d + f\sigma_{d+1}p^\lambda\partial_\mu\tau_\lambda\delta'(\tau_\rho p^\rho-m)\\
&= \partial_\mu f \sigma_d + f \bar\Gamma^\rho_{\rho\mu}\sigma_d + f\sigma_{d+1}p^\lambda\bar\Gamma^\sigma_{\lambda\mu}\tau_\sigma\delta'(\tau_\rho p^\rho-m) \\
&= \partial_\mu f \sigma_d + f\bar \Gamma^\rho_{\rho\mu}\sigma_d + f\sigma_{d+1}p^\lambda\bar\Gamma^\sigma_{\lambda\mu}\frac{\partial}{\partial p^\sigma}\delta(\tau_\rho p^\rho-m)\,.
\end{align}
Plugging it into Eq.~(\ref{eq:div_j_1st}), integrating the last term by parts, and comparing with Eq.~(\ref{eq:liouville}), shows that
\begin{equation}
    \bar\nabla_{\mu}\mathcal{I}^{\mu\alpha_2\cdots\alpha_n} = \frac{1}{m^{n-1}}\int  p^{\alpha_2} \cdots p^{\alpha_n} \left( X_L[f]+\frac{1}{2m^2}p^2H^{\mu}\,_{\rho}p^\rho\frac{\partial}{\partial p^\mu}f \right)\sigma_d\,.
    \label{eq:div_j_app}
\end{equation}
Integrating the last term by parts gives 
\begin{equation}
    \bar\nabla_{\mu}\mathcal{I}^{\mu\alpha_2\cdots\alpha_n} = \frac{1}{m^{n-1}}\int  p^{\alpha_2} \cdots p^{\alpha_n} X_L[f]\sigma_d -\frac{n-1}{2}H^{(\alpha_2}\,_{\rho}h_{\mu\nu}\mathcal{I}^{\mu\nu\rho|\alpha_3\cdots\alpha_n)}-H_{\mu\nu} v^\nu \mathcal{I}^{\mu\alpha_2\cdots \alpha_n}\,.
    \label{eq:div_j_app_2}
\end{equation}
Setting $v^\mu=u^\mu$ and $h_{\mu\nu}=g_{\mu\nu}$ gives Eq.~(\ref{eq:moments_covariant_0}).
\subsection{Hydrostatic conditions}
\label{app:killing_conditions}

Let us rephrase the Killing conditions in Eq.~(\ref{eq:killing_invariant}) using the Milne-invariant data defined in Eqs.~(\ref{eq:T-and-u}) and (\ref{eq:mu_inv}). Firstly,
\begin{align}
   \delta_{\chi}\tau_{\mu} &= \xi^{\nu}\partial_{\nu}\tau_{\mu}+\tau_{\nu}\partial_{\mu}\xi^{\nu}=\partial_\mu\left(\frac{1}{T}\right)-\frac{u^\nu}{T}H_{\mu\nu}=0~\implies\\
   &u^\mu\partial_\mu \left(\frac{1}{T}\right)=0 \,, \qquad h^{\mu\nu}\partial_\nu\left(\frac{1}{T}\right)  -H^\mu\,_{\nu} \frac{u^\nu}{T}=0\,.
   \label{app:killing_tau}   
\end{align}
Expanding $\partial_\mu(1/T)=-T^{-2}\partial_\mu T$ and multiplying by $T^2$ gives Eq.~(\ref{eq:killing_tau}). Secondly,
\begin{align}
   \delta_{\chi} A_{\mu} &= \xi^{\nu}\partial_{\nu}A_{\mu}+A_{\nu}\partial_{\mu}\xi^{\nu}+\partial_\mu \Lambda=\partial_\mu\left(\frac{\mu}{mT}\right)-\frac{u^\nu}{T}f_{\mu\nu}=0\,,\implies 
   \label{app:mu_hydro}\\
   & u^\mu\partial_\mu \left(\frac{\mu}{T}\right)=0 \,, \qquad h^{\mu\nu}\partial_\nu\left(\frac{\mu}{T}\right)  -mf^\mu\,_{\nu} \frac{u^\nu}{T}=0\,.
\end{align}
Using $u^\mu\partial_\mu \left(1/T\right)=0$ in the first equation and multiplying the last equation by $T$ gives Eq.~(\ref{eq:mu_hydro}). Finally, 
\begin{align}
   \delta_{\chi}h^{\mu\nu} = &\xi^{\lambda}\partial_{\lambda}h^{\mu\nu}-h^{\mu\lambda}\partial_{\lambda}\xi^{\nu}-h^{\nu\lambda}\partial_{\lambda}\xi^{\mu}\nn
   =&-h^{\mu\lambda}\bar \nabla_\lambda\xi^\nu - h^{\nu\lambda}\bar \nabla_\lambda\xi^\mu +u^\mu H^\nu{}_\lambda\xi^\lambda + u^\nu H^\mu{}_\lambda\xi^\lambda \nn
   =&-2 \bar\nabla^{(\mu}\left(\frac{u^{\nu)}}{T}\right)+2u^{(\mu}H^{\nu)}\,_\lambda \frac{u^\lambda}{T}=0\,.
   \label{app:killing_h}
\end{align}
Using Eq.~(\ref{app:killing_tau}) in Eq.~\eqref{app:killing_h} gives 
\begin{equation}
    \bar \nabla^{(\mu}u^{\nu)}=0\,.
\end{equation}
Taking the trace with $g_{\mu\nu}$ produces $\theta=\bar \nabla_\mu u^\mu=0$, while projecting out the traceless part produces $\sigma^{\mu\nu}=0$.
\subsection{Conservation of energy}
\label{app:energy}

By combining Eqs.~\eqref{eq:currents_define},~\eqref{eq:moments_covariant},~\eqref{eq:c_conditions} we obtain
\begin{equation}
    \left(\bar\nabla_\mu +H_{\mu\rho}u^\rho\right)\mathcal{E}^\mu = \frac{m}{2}\mathcal{I}^{\gamma\delta\mu}\bar\nabla_\mu g_{\gamma\delta}-\frac{m}{2}g_{\mu\nu}H^{(\mu}\,_{\rho}g_{\beta\gamma}\mathcal{I}^{\beta\gamma\rho|\nu)}\,.
\end{equation}
Furthermore,
\begin{equation}
\begin{split}
    \bar\nabla_{\mu}g_{\gamma\delta} & = \left(\bar P^\alpha_\gamma+\tau_\gamma u^\alpha\right)\bar\nabla_{\mu}g_{\alpha \delta} = g_{\gamma\beta}\bar\nabla_{\mu}\left(h^{\alpha\beta}g_{\alpha\delta}\right)-\tau_\gamma g_{\alpha\delta}\bar\nabla_\mu u^\alpha \\
    &= g_{\gamma\beta}\bar\nabla_{\mu}\left(\delta^\beta_\delta - \tau_\delta u^\beta\right)-\tau_\gamma g_{\alpha\delta}\bar\nabla_\mu u^\alpha = -2 \tau_{(\gamma}g_{\alpha\delta)}\bar\nabla_\mu u^\alpha\,
\end{split} \label{eq:div_V_fromC}
\end{equation}
and
\begin{equation}
\begin{split}
    -\frac{m}{2}g_{\mu\nu} H^{\mu}\,_{\rho}g_{\beta\gamma}\mathcal{I}^{\beta\gamma\rho\nu}&=-\frac{m}{2}\left(\delta^\mu_\nu - u^\mu \tau_\nu\right)H_{\mu\rho}g_{\beta\gamma}\mathcal{I}^{\beta\gamma\rho\nu} \\
    &= - \frac{m}{2}H_{\nu\rho}g_{\beta\gamma}\mathcal{I}^{\beta\gamma\rho\nu}+\frac{m}{2}u^\mu H_{\mu\rho}g_{\beta\gamma}\mathcal{I}^{\beta\gamma\rho} \\
    &= -H_{\rho \mu} u^\mu \mathcal{E}^\rho\,.
\end{split}
\end{equation}
Altogether, we obtain
\begin{equation}
    \left(\bar\nabla_\mu +H_{\mu\rho}u^\rho\right)\mathcal{E}^\mu = -H_{\mu\rho}u^\rho\mathcal{E}^\mu-\mathcal{T}^{\mu\nu} g_{\rho(\mu}\bar\nabla_{\nu)}u^\rho \,.
\end{equation}

\subsection{Derivation of the equations of motion (\ref{eq:deriv2_fifth})}
\label{app:deriv_eom}

In this subsection, we derive the equations of motion in Eq.~\eqref{eq:deriv2_fifth} from the evolution of the $k$-th moment of the distribution function. Shifting the projection operator inside the covariant derivative in Eq.~\eqref{eq:deriv2_init}, we get
\begin{align}
    \nabla_\mu\lb \bar{P}^{\alpha_1}_{\alpha_1'}\cdots \bar{P}^{\alpha_k}_{\alpha_k'} \mathcal{I}^{\mu\alpha_1'\cdots\alpha_k'}\rb -  \nabla_\mu \lb \bar{P}^{\alpha_1}_{\alpha_1'}\cdots \bar{P}^{\alpha_k}_{\alpha_k'} \rb\mathcal{I}^{\mu\alpha_1'\cdots\alpha_k'}=K^{\alpha_1\cdots\alpha_k}\,.
    \label{eq:deriv2_init1}
\end{align}
The covariant derivative of the projection operator is
\begin{align}
    \nabla_\mu \bar{P}^\alpha_{\alpha'} & = \nabla_\mu\left(\delta^\alpha_{\alpha'}-u^\alpha \tau_{\alpha'}\right)=-\tau_{\alpha'} \nabla_\mu u^\alpha
    \nn
    & =-\tau_{\alpha'}\left(\bar{P}^\alpha_\beta+\tau_\beta u^\alpha\right)\nabla_\mu u^\beta=-\tau_{\alpha'}\bar{P}^\alpha_\beta \nabla_\mu u^\beta\,.
    \label{eq:deriv-of-projec-opera}
\end{align}
Using Eq.~\eqref{eq:deriv-of-projec-opera} in Eq.~\eqref{eq:deriv2_init1} produces
\begin{align}
    &\nabla_\mu \lb  \bar{P}^{\alpha_1}_{\alpha_1'}\cdots \bar{P}^{\alpha_k}_{\alpha_k'} \mathcal{I}^{\mu\alpha_1'\cdots\alpha_k'}\rb  
    \nn
    &+\left(\bar{P}^{\alpha_1}_{\alpha_1'}\cdots \bar{P}^{\alpha_{k-1}}_{\alpha_{k-1}'} \tau_{\alpha_k'}h^{\alpha_k\sigma}+\cdots+\tau_{\alpha_1'}h^{\alpha_1\sigma}\bar{P}^{\alpha_2}_{\alpha_2'}\cdots \bar{P}^{\alpha_k}_{\alpha_k'}\right)\left(g_{\lambda\sigma}\nabla_\mu u^\lambda\right)\mathcal{I}^{\mu\alpha_1'\cdots\alpha_k'} 
    = K^{\alpha_1\cdots\alpha_k}\,. 
    \label{eq:deriv2_second}
\end{align}
Furthermore, we notice the equality
\begin{equation}    
    \bar{P}^{\alpha_1}_{\alpha_1'}\cdots \bar{P}^{\alpha_k}_{\alpha_k'} \mathcal{I}^{\mu\alpha_1'\cdots\alpha_k'} = u^\mu I^{\alpha_1\cdots\alpha_k}+I^{\mu\alpha_1\cdots\alpha_k}\,,
    \label{eq:pro-I-in-split}
\end{equation}
so that Eq.~\eqref{eq:deriv2_second} can be written as
\begin{align}
    \nabla_\mu \left(u^\mu I^{\alpha_1\cdots\alpha_k}\right)&+\nabla_\mu I^{\mu\alpha_1\cdots\alpha_k} 
    +\big[h^{\alpha_k\sigma}\left(u^\mu I^{\alpha_1\cdots\alpha_{k-1}}+I^{\mu\alpha_1\cdots\alpha_{k-1}}\right)+\cdots
    \nn
     &\cdots +h^{\alpha_1\sigma}\left(u^\mu I^{\alpha_2\cdots \alpha_k}+I^{\mu\alpha_2\cdots \alpha_k}\right)\big]\left(g_{\lambda\sigma}\nabla_\mu u^\lambda\right)
    = K^{\alpha_1\cdots\alpha_k}\,.
    \label{eq:deriv2_fourth}
\end{align}
The Lie derivative of $I^{\alpha_1\cdots\alpha_k}$ with respect to $u^\mu$ is
\begin{equation}
\begin{split}
    \pounds_u I^{\alpha_1\cdots\alpha_k} &= u^\mu \partial_\mu I^{\alpha_1\cdots\alpha_k} - I^{\mu\alpha_2\cdots\alpha_k}\partial_\mu u^{\alpha_1} - \cdots - I^{\alpha_1\cdots\alpha_{k-1}\mu}\partial_\mu u^{\alpha_k}\\
    & = u^\mu \nabla_\mu I^{\alpha_1\cdots\alpha_k} - I^{\mu\alpha_2\cdots\alpha_k}\nabla_\mu u^{\alpha_1} - \cdots - I^{\alpha_1\cdots\alpha_{k-1}\mu}\nabla_\mu u^{\alpha_k}\,,
\end{split}
\end{equation}
so that Eq.~(\ref{eq:deriv2_fourth}) can be written as
\begin{align}
    \left(\pounds_u+\nabla_\mu u^\mu\right) I^{\alpha_1\cdots \alpha_k}&+\nabla_\mu I^{\mu\alpha_1\cdots\alpha_k} 
    +\big[h^{\alpha_k\sigma}\left(u^\mu I^{\alpha_1\cdots\alpha_{k-1}}+2I^{\mu\alpha_1\cdots\alpha_{k-1}}\right)+\cdots
    \nn
     &\cdots +h^{\alpha_1\sigma}\left(u^\mu I^{\alpha_2\cdots \alpha_k}+2I^{\mu\alpha_2\cdots \alpha_k}\right)\big]\left(g_{\lambda\sigma}\nabla_\mu u^\lambda\right)
    = K^{\alpha_1\cdots\alpha_k}\,.
\end{align}


\subsection{Evaluating \texorpdfstring{$S^{\alpha\beta}$}{TEXT} and \texorpdfstring{$S^{\alpha\beta\gamma}$}{TEXT}}
\label{app:deriv_1storder}

The tensors $S^{\alpha_1\alpha_2\ldots\alpha_k}$ are defined through Eqs.~(\ref{eq:F_covariant}) and (\ref{eq:I_C_defs}) as the spatial parts of the $k$-th moments of $X_L[f_0]$. For that reason we start this section by expressing $X_L[f_0]$ in a form convenient for later manipulations.

In the definition of the hydrodynamic distribution function $f_0$, we replace the chemical potential $\mu$ with the number density $n$ with the use of Eq.~\eqref{eq:current_ideal}: 
\begin{equation}    
    f_0 =c^{-1} n\lb\frac{1}{2\pi m T}\rb^{d/2} \exp\left[-\frac{1}{2mT}g_{\alpha\beta}p^\alpha p^\beta\right]\,.
\end{equation}
We have also used the Milne-invariance of $f_0$ and the fact that $g_{\mu\nu}u^\nu=0$. Now we apply the Liouville operator in Eq.~\eqref{eq:liouville} to the equilibrium Boltzmann distribution functions, getting
\begin{equation}
\begin{split}
    X_L[f_0]= &\frac{p^\mu}{m} \partial_\mu f_0-\frac{1}{m}\Gamma^\mu_{\gamma\delta}p^\gamma p^\delta \frac{\partial f_0}{\partial p^\mu} \\
     = &-\frac{p^\mu}{m}\left[\frac{1}{2mT}p^\gamma p^\delta\partial_\mu g_{\gamma\delta} -\frac{1}{2mT^2}\lb g_{\gamma\delta}p^\gamma p^\delta-dmT\rb\partial_\mu T - \frac{\partial_\mu n}{n}\right]f_0+ \frac{1}{mT}\Gamma^\mu_{\gamma\delta}p^\gamma p^\delta g_{\mu\rho}p^\rho f_0
    \\
    =& -\frac{p^\mu}{m}\left[\frac{1}{2mT}p^\gamma p^\delta\nabla_\mu g_{\gamma\delta}-\frac{1}{2mT^2}\lb g_{\gamma\delta}p^\gamma p^\delta-dmT\rb\partial_\mu T- \frac{\partial_\mu n}{n}\right]f_0\,.
    \label{eq:deriv4_1}
\end{split}
\end{equation}
In the last expression, we have replaced the ordinary derivative with the covariant derivative using the relation
\begin{align}
    \partial_\mu g_{\gamma\delta} = \nabla_\mu g_{\gamma\delta} + \Gamma^\alpha_{\gamma\mu}g_{\alpha\delta}+\Gamma^\alpha_{\delta\mu}g_{\alpha\gamma}\,.
\end{align}
From Eq.~\eqref{eq:div_V_fromC} follows that $p^\gamma p^\delta \nabla_\mu g_{\gamma\delta}=-2mp^\gamma g_{\gamma \delta}\nabla_\mu u^\delta$. Eq.~\eqref{eq:deriv4_1} can be rewritten as follows:
\begin{equation}
\begin{split}
    X_L[f_0]
     =&  \frac{\lb p^\mu-mu^\mu\rb}{m}\left[  \frac{\partial_\mu n}{n} +\frac{1}{T}g_{\gamma\delta}p^\gamma\nabla_\mu u^\delta +\frac{1}{2mT^2}\lb g_{\gamma\delta}p^\gamma p^\delta-dmT\rb\partial_\mu T \right]f_0\\
    &+u^\mu\left[  \frac{\partial_\mu n}{n} +\frac{1}{T}g_{\gamma\delta}p^\gamma\nabla_\mu u^\delta +\frac{1}{2mT^2}\lb g_{\gamma\delta}p^\gamma p^\delta-dmT\rb\partial_\mu T \right]f_0\,.
    \label{eq:deriv4_3}
\end{split}
\end{equation}
Using the equations of motion in Eqs.~\eqref{eq:eoms_thermo_covariant-a}-\eqref{eq:eoms_thermo_covariant-c}, we can eliminate the terms containing $u^\mu \partial_\mu n$, $u^\mu \nabla_\mu u^\delta$, and $u^\mu \partial_\mu T$. This way we obtain 
\begin{align}
    X_L[f_0] = & \left[\frac{1}{mT}g_{\gamma\alpha}p^\gamma\left(p^\mu-mu^\mu\right)\left(\nabla_{\mu}u^\alpha-\frac{1}{d}\delta^\alpha_\mu\nabla_{\nu}u^\nu\right)\right.\nn
    & \left. +\frac{1}{2m^2T^2}\left\{ g_{\gamma\alpha}p^\gamma p^\alpha-(d+2)mT\right\} (p^\mu-mu^\mu)\partial_\mu T \right. \nn
    &\left.  - \frac{1}{dmnT^2}\lb g_{\gamma\alpha}p^\gamma p^\alpha-dmT\rb \lb\tau^{\mu\nu}g_{\rho\mu}\nabla_{\nu}u^\rho+\nabla_\mu q^\mu\rb
     -\frac{1}{mnT}g_{\gamma\delta}p^\gamma\nabla_\mu \tau^{\mu\delta}\right]f_0\,.
\label{eq:Lx_fid}
\end{align}

To derive the first-order constitutive relations, we drop the higher-order gradient correction in Eq.~\eqref{eq:Lx_fid} and write
\begin{align}
    X_L[f_0] = & \left[\frac{1}{mT}g_{\gamma\alpha} p^\gamma\left(p^\mu-mu^\mu\right)\left(\nabla_{\mu}u^\alpha-\frac{1}{d}\delta^\alpha_\mu\nabla_{\nu}u^\nu\right)\right. \nn
    & \left. +\frac{1}{2m^2T^2}\left\{ g_{\gamma\alpha}p^\gamma p^\alpha-(d+2)mT\right\} (p^\mu-mu^\mu)\partial_\mu T \right]f_0\,.
    \label{eq:Lx_fid_1storder}
\end{align}
Written in this form, the formula for $X_L[f_0]$ resembles the one familiar from the textbook approach to the Chapman-Enskog expansion \cite{chapman1990mathematical,lifshitz1981physical,dorfman2021contemporary}. Now we proceed to calculate the moments of $X_L[f_0]$ one by one. The zeroth moment is
\begin{equation}
    S = -\int \sigma_{d+1}\delta(\tau_\rho p^\rho-m) X_L[f_0] = 0\,,
    \label{eq:deriv5_1}
\end{equation}
the integral easily computed after changing the variables $p^\mu \rightarrow (p^\mu+mu^\mu)$. The first moment is
\begin{align}
    S^\alpha & =-\frac{1}{m}\bar P^\alpha_{\alpha'}\int \sigma_{d+1}\delta(\tau_\rho p^\rho-m) p^{\alpha'} X_L[f_0]=0\,,
    \label{eq:deriv5_2}
\end{align}
where we also use the change of variables  $p^\mu \rightarrow (p^\mu+mu^\mu)$ and use Eq.~\eqref{eq:deriv5_1}. Similarly, using the zeroth and first moment, we calculate the second moment as 
\begin{equation}
    S^{\alpha\beta}
     = - \frac{2 n T}{m} h^{\rho \langle \alpha }\nabla_\rho u^{\beta\rangle} = -\frac{2 n T}{m}\sigma^{\alpha\beta}\,.
     \label{eq:deriv5_3}
\end{equation}
Finally, we get the third moment as  
\begin{equation}
\begin{split}
    S^{\alpha\beta\gamma} &= -\frac{3nT}{m^2}h^{(\alpha\beta}\partial^{\gamma)} T\,.
\end{split} \label{eq:deriv5_4}
\end{equation}

\bibliography{biblio}
\bibliographystyle{jhep}

\end{document}